\documentclass[12pt]{article}
\usepackage{amsfonts}
\usepackage{amssymb}
\usepackage{amsmath}
\usepackage{amsthm}
\usepackage{color}
\usepackage{multirow}
\usepackage[unicode=true,pdfusetitle, bookmarks=true,bookmarksnumbered=false,bookmarksopen=false, breaklinks=false,pdfborder={0 0 1},backref=section,colorlinks=true, linkcolor = blue, citecolor = blue]{hyperref}

\usepackage{graphicx}

\usepackage{bbm}

\usepackage{natbib}
\usepackage{todonotes}

\usepackage{url}

\hypersetup{colorlinks=true, linkcolor=blue}

\newtheorem{theorem}{{\bf \sc Theorem}}

\newtheorem{corollary}{{\bf \sc Corollary}}
\newtheorem{proposition}{{\bf \sc Proposition}}
\newtheorem{remark}{{\bf \sc Remark}}

\newtheorem{definition}{{\bf \sc Definition}}

\newcommand{\pidefer}{\ensuremath{\Pi^{{\textrm{defer}}}}}
\newcommand{\piadopt}{\ensuremath{\Pi^{{\textrm{adopt}}}}}
\newcommand{\mc}[1]{\ensuremath{\mathcal{#1}}}

\newcommand{\delrj}[1]{}

\newtheorem{assumption}{{\bf \sc Assumption}}
\newtheorem{mfassumption}{{\bf \sc Mean-Field Assumption}}

\begin{document}


\title{Pricing and Referrals in Diffusion on Networks}


%
%
\author{Matt V. Leduc%
\thanks{(Corresponding author). Dept. of Management Science \& Engineering. Leduc is also a research scholar at IIASA (Austria). Email: mattvleduc@gmail.com. }%
\and Matthew O. Jackson%
\thanks{Department of Economics, Stanford University. Jackson is also a fellow of CIFAR, and an external faculty member of the Santa Fe Institute. Email: jacksonm@stanford.edu.}%
\and Ramesh Johari%
\thanks{Dept. of Management Science \& Engineering. Email: ramesh.johari@stanford.edu.}%
}
%

\date{Stanford University \\Draft: June 2017 \\ \ \\ \small{For the version of this article published in \textit{Games and Economic Behavior}, the reader is referred to https://doi.org/10.1016/j.geb.2017.05.011}}

%
%

\maketitle

\begin{abstract}

When a new product or technology is introduced, potential consumers can learn its quality by trying it, at a risk, or by letting others try it and free-riding on the information that they generate. We propose a dynamic game to study the adoption of technologies of uncertain value, when agents are connected by a network and a monopolist seller chooses a profit-maximizing policy. Consumers with low degree (few friends) have incentives to adopt early, while consumers with high degree have incentives to free ride. The seller can induce high-degree consumers to adopt early by offering referral incentives - rewards to early adopters whose friends buy in the second period. Referral incentives thus lead to a `double-threshold strategy' by which low and high-degree agents adopt the product early while middle-degree agents wait. We show that referral incentives are optimal on certain networks while inter-temporal price discrimination is optimal on others.

\noindent {\bf Keywords}: Network Games, Technology Adoption, Social Learning, Word-of-Mouth, Network Diffusion, Dynamic Pricing, Referral Incentives.

\noindent {\bf JEL Codes}: D85, C72, L11, L12
\end{abstract}

\thispagestyle{empty}

\setcounter{page}{0} \newpage

\section{Introduction}

In this paper, we study the interplay between social learning,
efficient product diffusion, and the optimal pricing policy of a monopolist.
More precisely, we study the adoption dynamics of a technology of uncertain value,
when {forward-looking} agents interact through a network and must decide not
only \em whether \em to adopt a new product, but also \em when \em to adopt it. Uncertainty leads to informational free-riding: a potential consumer may wish to delay adoption in order to let other agents bear the risks of experimenting with the technology and learn from their experiences.   This complicates the problem of technology adoption and can lead to inefficiencies in diffusion processes, as there are risks from being an early adopter and externalities in early adoption decisions.
The possibility of free-riding induces a specific form of social inefficiency:
agents with relatively few friends (low degree) have the greatest incentives to try
the product since they have the least opportunity to observe others' choices.  Given
the risks of experimentation, it would be more socially efficient to have high-degree
agents experiment since they are observed by many others, thus lowering the number of
experimenters needed to achieve a given level of information in the society.
This problem occurs in many settings: not only do consumers benefit from the
research of friends and relatives into new products, but farmers benefit from
the experience of other farmers with a new crop. Likewise in industry,
research spills over to other firms. People benefit from the experience of their
friends and relatives regarding a vaccine of unknown side effects. In developing
countries, villagers may learn about a new program from the experience of
other members of the community.

We study this problem in a two-period network game in which a monopolist (or social planner)
can induce people to experiment with the product in the first period via
two types of incentives: price discounts and referral rewards (payments to an agent who tries the product early based on how many of that agent's friends later adopt the product).\footnote{Referral rewards are seen in many settings
with new products or technologies.   For instance, in July of 2015, Tesla Motors announced a program by which an owner of a Model S
Sedan would receive a 1000 dollar benefit if the owner referred a friend who also
buys a Model S Sedan (Bloomberg Business News, ``Musk Takes Page From
PayPal With Tesla Referral Incentive,'' August 31, 2015).
Dropbox rapidly grew from around one
hundred thousand users in the fall of 2008 to over four million by the spring of
2010, with more than a third of the signups coming through its official referral
program that offered free storage to both referrer and referree (Forbes, ``Learn The Growth Strategy That Helped Airbnb And Dropbox Build Billion-Dollar Businesses,'', Feb. 15, 2015).  Such programs have been used by many new companies from Airbnb to Uber, and also by large existing companies when introducing new products (e.g., Amazon's Prime). }
Price discounts induce more agents to try the product early, but are biased towards low-degree agents since they are the ones with the greatest incentives to try early in any case.  In contrast, referral rewards induce
high-degree agents to try the product early since they have more friends to refer in the second period and thus expect greater referral rewards.
We show that if sufficient referral incentives are in place then early adoption is characterized by a double-threshold pattern in which both low and high-degree agents adopt early while middle-degree agents choose to delay adoption and learn from the behavior of others, before making later adoption decisions.  The specifics of
the lower and upper thresholds depend on the combination of prices and referral incentives.
We then study a monopolist's optimal pricing strategy.  The monopolist's incentives are partly aligned with social efficiency since it is costly to induce first-period experimentation - either price discounts or referral incentives must be offered and the monopolist would like to minimize such payments and maximize the number of eventually informed high-paying adopters.  The optimal strategy, however, depends on network structure via the relative numbers of agents of different degrees.
We characterize the optimal policies for some tractable degree distributions and provide insights into the more general problem.
A rough intuition is that if the network is fairly regular, then referral incentives
are less effective and price discounts are the main tool to maximize profits.
If instead, there is sufficient heterogeneity in the degree distribution and there are
some agents of sufficiently high degree, then referral incentives are more
profitable.   In some limiting cases, in which the network has high enough degree,
referral incentive policies (with no price discounts) are both profit maximizing and socially efficient.

Our approach enriches an early literature on social learning (e.g., \cite{Chamley94}, \cite{Chamley04}, \cite{Gul95} and \cite{Rogers05}) that focused on delayed information collection through stopping games. Our analysis brings in the richer network setting and analyzes a monopolist's pricing problem.
Our network modeling builds on the growing literature on network diffusion,\footnote{See \cite{JacksonYarivHandbook} for a recent review of the field, and \cite{GoelWattsEC12} and \cite{LeskovecPredict} for recent empirical work.} and uses the mean-field approach to study diffusion developed in \cite{JacksonYariv05,JacksonYariv,Manshadi09,Galeotti10, MLeducThesis,leduc2015protection}.
Our paper is also related to a recent literature modeling monopolistic marketing in social networks\footnote{\cite{papanastasiou2016dynamic} study dynamic pricing in the presence of social learning and free-riding, but without a network structure.} (e.g.,  \cite{hartline,Candogan12,Bloch13,fainmesser,saaskilahti,ShinMonopPricing}) that builds on an earlier literature of pricing with network effects (\cite{farrell,katz}).
Our approach differs as it considers the dynamic learning in the network about
product quality, rather than other forms of complementarities, and works off of
inter-temporal price discrimination that derives from network structure and information flows.\footnote{There are some papers
that have looked explicitly at the dynamics of adoption and marketing, such as \cite{hartline}, but again based on other complementarities and the complexities of computing an optimal strategy rather than dynamic price discrimination in the face of social learning.}
This enriches an earlier literature on price discrimination that focuses mainly on information gathering costs and heterogeneity in consumers' tastes or costs of information acquisition and/or demand uncertainty for the monopolist (\cite{Kalish,Lewis,Courty,Dana,Bar-Isaac,Nockea}).
Thus, our approach is quite complementary, as it not only applies to different settings but it is also based on a different intuition: the pricing policy in our case is used
as a screening device on agents' network characteristics. The monopolist does not observe the network but instead induces agents with certain network characteristics to experiment with the product and potentially later induce other agents to also use it. The latter can then be charged different prices.  Referral incentives are useful because they induce highly-connected individuals to adopt early and thus take advantage of their popularity, solving an informational inefficiency at the same time as increasing profits.

The paper is organized as follows. Section \ref{SecModel} presents the dynamic network game in a finite setting. Payoffs are defined and basic assumptions are stated.
Section \ref{sec:MF} develops the mean-field equilibrium framework that allows us
to study the endogenous adoption timing in a tractable way while imposing a realistic
cognitive burden on agents. Section \ref{SecDynamicPricing} illustrates how the
dynamic game allows us to study a large class of dynamic pricing policies. Section \ref{sec:profitMax} studies the monopolist's profit maximization problem. Policies
involving referral incentives and policies using inter-temporal price discrimination
are compared. Section \ref{Conclusion} concludes. For clarity of exposure, all proofs
are presented in an appendix.

\section{The Dynamic Adoption Game With Finitely Many Agents}
\label{SecModel}

We first analyze the equilibrium of a dynamic adoption game just among a collection of agents, fixing a prior belief about the quality of the product.  We come back to analyze the pricing strategies
of a monopolist and derivation of agents' beliefs later.

A set of agents are interconnected through a network that we approximate with a mean-field model, described in detail below.
Each agent (consumer) $i$ learns her {\em own} degree $d_i$, but not the degrees of her neighbors.

Two periods are denoted by times, $t\in \{0,1\}$.

Agent $i$ can choose to adopt at time $t = 0$, or adopt at time $t = 1$, or not to adopt at all. If the agent adopts at $t=0$, she can choose to discontinue the use of the technology at $t=1$.

We let $X_{i,t}$ denote the number of $i$'s neighbors who adopt at time $t$.  

The technology is of either {\em High} or {\em Low} quality, depending on an unknown state variable (the {\em quality}) $\theta \in \{H, L\}$.
Let
$p$ denote the probability that $\theta=H$.
If the technology is of high quality ($\theta = H$), its value at $t=0$ is $A_0^H > 0$ and its value at $t=1$ is $A_1^H > 0$.  If the technology is of low quality ($\theta = L$), then its value at $t=0$ is $A_0^L < 0$ and its value at $t=1$ is $A_1^L < 0$. Agents have a common prior belief $p \in (0,1)$ that $\theta = H$.

A key informational assumption is that {\em if any neighbor of agent $i$ adopts at $t = 0$, then agent $i$ learns the quality of the good prior to choosing her action at $t = 1$.}  This assumption enables social learning via free-riding.  In particular, if agent $i$ adopts early, he ``teaches'' his neighbors the quality\footnote{This is similar to a word-of-mouth process, as studied in \cite{campbell2013word}.}.

An agent who adopts in the first period (at $t=0$) earns a {\em referral payment}
$\eta \geq 0$ for each neighbor who adopts after him (at $t=1$).
This can take different forms. For instance, it can be an altruistic
benefit derived from helping a friend.
It can also be a payment received from the seller, as will be the case
in our analysis of the monopolist's problem, or it might be the sum of such terms.
We stress that an agent receives $\eta$ for each neighbor who adopts
after him even if this neighbor is also connected to other early adopters.
In practice, it may be that for each late adopter, only one early adopter
gets a referral reward.  An extension in which the referral payment is given
only to a randomly-selected early adopter would not change the qualitative nature of our results,
although
it complicates the calculations.

To introduce the basic structure and illustrate some of the
incentives, we first normalize prices to zero in both periods.
We analyze the case with prices further below.

Any discounting of payoffs are captured in the values $A_1^\theta$.  Table \ref{tab:payoffs} summarizes agent $i$'s payoffs for using the technology at different times as a function of the state.
\begin{table}[h]
\begin{center}
  \begin{tabular}{ |l | c | r| }
  \hline
       & $\theta = H$ & $\theta = L$ \ \ \ \  \\ \hline
    \textrm{}\ t = 0 & $A_0^H + \eta X_{i,1}$ & $A_0^L + \eta X_{i,1}$ \\ \hline
    \textrm{}\ t = 1 & $A_1^H $ & $A_1^L  \ \ \ \ \ $ \\
    \hline
  \end{tabular}
  \caption[Payoffs for Use at Different Times]{Payoffs for Use at Different Times}\label{tab:payoffs}
  \end{center}
  \end{table}

The following is assumed about payoffs to focus on the nontrivial case in which learning is valuable.
\begin{assumption}
\label{as:params}
$pA^H_1 + (1-p)A^L_1 < 0$.
\end{assumption}

This assumption implies that if agent $i$ does not learn the quality of the technology by time $t = 1$, she will not adopt.  The fact that $A^H_1>0$ and $A^L_1<0$, on the other hand, ensures that if agent $i$ learns the quality of the technology by time $t = 1$, then she will adopt if $\theta = H$ and not adopt if $\theta = L$.  Under this information and payoff structure, the time $1$ decision problem of an agent who has not already adopted is simplified.  We summarize this in the following remark.

\begin{remark}
\label{rem:t=1}
Suppose an agent $i$ has not adopted at $t = 0$.  Then if $X_{i,0} > 0$, agent $i$ adopts at $t = 1$ if $\theta = H$, and does not adopt if $\theta = L$.  If $X_{i,0} = 0$, agent $i$ does not adopt at $t = 1$.
\end{remark}

Thus, we may rewrite the payoff table as a function of an agent $i$'s strategy and whether an agent's neighbors adopt as follows,
presuming that the agent follows the optimal strategy in the second period:
\begin{table}[h]
\begin{center}
  \begin{tabular}{ |l | c | r| }
  \hline
       & $\theta = H$ & $\theta = L$ \ \ \ \  \\ \hline
    Adopt at $t = 0$ & $A_0^H +A_1^H+ \eta X_{i,1}$ & $A_0^L + \eta X_{i,1}$ \\ \hline
    Not adopt at $t = 0$ and $X_{i,0} > 0$ & $A_1^H $ & $0   \ \ \ \ \ $ \\ \hline
    Not adopt at $t = 0$ and $X_{i,0} = 0$ & $0 $ & $0  \ \ \ \ \ $ \\
    \hline
  \end{tabular}
  \caption[Total Expected Payoff Based on $t=0$ Adoption Decision]{Total Expected Payoff Based on $t=0$ Adoption Decision}
  \end{center}
  \end{table}

We also make an assumption that it is not in an agent's interest to adopt at time 0 unless there is a sufficient referral incentive or a sufficient option value of learning.  The assumption is that the time 0 expected payoff, just considered in isolation, is negative:
\begin{assumption}
\label{as:AlwaysFreeriding}
$\bar{A}:=p (A_0^H+A_1^H) + (1 - p)A_0^L < pA^H_1$, or $p A_0^H+ (1 - p)A_0^L < 0$.
\end{assumption}
In the absence of this assumption, all agents prefer (irrespective of referral rewards)
to adopt in the first period as they would have a positive expectation in the first period, plus the benefit of learning as well as any referral rewards, and so the problem becomes uninteresting.\footnote{Even the monopoly pricing problem becomes relatively uninteresting, as the monopolist (being sure of the high value) can extract the full first period value and second period value, without any referral incentives.}

In principle, we would be interested in searching for a {perfect Bayesian equilibrium} of this game.
While the previous remark simplifies the time $1$ decision problem of an agent, the time $0$ problem is
intractable as a function of the graph: each agent's decision depends on a forecast of the agent's neighbors' strategies, which
depend on their forecasts, etc., and all of these correlate under the distribution from which the graph is drawn.    This is a common problem in such network games; see, e.g., \cite{JacksonYariv05,JacksonYariv,Manshadi09,Galeotti10,AdlakhaJohariWeintraub10} for related models that face similar issues.  In the next section, we present a mean-field model with which to study this game.

\section{A Mean-Field Version of the Game}
\label{sec:MF}

We now describe an approximation of the dynamic adoption game that is more tractable to solve: the mean-field game.  This can be thought of symmetrizing the model so that agents have similar beliefs about their neighbors' behaviors, which makes the game tractable and holds at the limit for large networks.  We examine the accuracy of the approximation in the appendix for a couple of special cases for which calculations are feasible.\footnote{The mean-field equilibrium is a variation on that in
\cite{Galeotti10}.}

A {mean-field equilibrium} for our dynamic game captures the standard notion that agents best respond to the distribution induced by their actions.

The {\em degree distribution}\footnote{Throughout the paper, we use the phrase \textit{degree distribution} to indicate the density function of the degree. When referring to the \textit{cumulative distribution function (CDF)}, we will do so explicitly.} is captured by the probability, $f(d)$ for $d \geq 1$, that any given node
had degree $d$. Note that agents of degree $0$ face trivial decisions (since they do not play a game) and so we ignore those nodes, and normalize the degree
distribution so that $f(0)=0$.

Agents reason about the graph structure in a simple way through the degree distribution:\footnote{This
assumption is stronger than needed.  Correlation in degrees can also be introduced, with some
care to make sure that it is not too assortative, and so preserves the incentive structure in what follows.}
\begin{mfassumption}
\label{mfa:degree}
Each agent conjectures that the degrees of his or her neighbors are drawn i.i.d. according to the edge-perspective degree distribution $\tilde{f}(d) = \frac{f(d) d}{\sum_{d' } f(d') d'}$.
\end{mfassumption}

Our equilibrium concept is general enough to allow us to posit any belief over the degree of an agent's neighbor.\footnote{For more on this see
\cite{Galeotti10}.} To keep things concrete, we consider $\tilde{f}(d)$ since it is consistent with the degree distribution $f(d)$. However, $\tilde{f}(d)$ could technically be replaced by any belief over a neighbor's interaction level.

A second mean-field assumption addresses how an agent reasons about the adoption behavior of his neighbors:
\begin{mfassumption}
\label{mfa:adopt}
Each agent $i$ conjectures that each of his or her neighbors adopts at $t = 0$ with probability $\alpha$, independently across neighbors: in particular, an agent $i$ with degree $d$ conjectures that the number of neighbors $X_{i,0}$ that adopt at $t = 0$ is a Binomial($d, \alpha$) random variable.
\end{mfassumption}

In equilibrium, $\alpha$ matches the expected behavior of a random draw from the population according to the edge-perspective degree distribution.

\subsection{Best Responses at $t = 0$}
\label{sec:BestResponse}

Given the mean-field structure, what strategy should an agent follow as a function of $\alpha$?

Note that for any realization of $X_{i,0}$, an agent $i$'s best response at $t=1$ is characterized by Remark \ref{rem:t=1}.  Thus, we focus on an agent's optimal strategy at $t = 0$.

\begin{definition}
A {\em mean-field strategy} $\mu: \mathbbm{N}_+ \rightarrow [0,1]$ specifies, for every $d >0 $, the probability that an agent of degree $d$ adopts at $t = 0$. We denote by $\mc{M}$ the set of all mean-field strategies.
\end{definition}


The expected payoff of an agent of degree $d$ who adopts at $t = 0$ as a function of $\alpha$ is given by:
\begin{equation}
\label{eq:mf_adopt_t0}
\piadopt(\alpha, d) =
p (A_0^H+A_1^H) + (1 - p)A_0^L + \eta p (1 - \alpha) d.
\end{equation}
This expression is derived as follows.  The first and second terms are the direct expected payoff this agent earns from adopting the technology
under prior $p$. The third term is the referral incentive of $\eta$ for any neighbors who adopt at $t = 1$ and who did not adopt at $t = 0$; there are an expected $d(1-\alpha)$ such neighbors who might adopt.  From Remark \ref{rem:t=1}, such a neighbor adopts if and only if the quality is good, i.e., $\theta = H$; and this event has probability $p$.

On the other hand, what is the payoff of such an agent if she does not adopt at $t = 0$?  In this case, the given agent only adopts at $t = 1$ if at least one of her neighbors adopts, {\em and} the technology is good.  This expected payoff is:
\begin{equation}
\label{eq:mf_delay_t0}
\pidefer(\alpha, d) = p A^H_1 (1 - (1-\alpha)^d).
\end{equation}

Obviously, an agent strictly prefers to adopt early if
$\piadopt(\alpha, d)-\pidefer(\alpha, d)>0$; strictly prefers not to if
$\piadopt(\alpha, d)-\pidefer(\alpha, d)<0$; and is indifferent (and willing to mix) when the two expected utilities are equal.

Let $\mc{BR}_d(\alpha) \subset [0,1]$ denote the set of best responses for a degree $d$ agent given $\alpha$.
Let $\mc{BR}(\alpha) \subset \mc{M}$ denote the space of mean-field best responses given $\alpha$; i.e.,
\[ \mc{BR}(\alpha) = \prod_{d \geq 1} \mc{BR}_d(\alpha). \]

\subsection{Mean-Field Equilibrium}

We now define an equilibrium.

Given $\mu$, the probability that any given neighbor adopts in the first period, now denoted as a function $\alpha(\mu)$, is determined as follows:

\begin{equation}
 \alpha(\mu) = \sum_{d \geq 1} \tilde{f}(d) \mu(d).
\label{eq:alpha}
\end{equation}

Note that $\alpha(\mu)$ lies in [0,1], since $\mu(d)\in [0,1]$.

Since $\alpha(\mu)$ corresponds to the probability that any agent will be informed of the quality of the technology by a randomly-picked neighbor, it has a natural interpretation as an agent's ``social information.''  For this reason we refer to $\alpha(\mu)$ as the {\em informational access} offered by the strategy $\mu$.

\begin{definition}[Mean-field equilibrium]
\label{def:mfe}
A mean-field strategy $\mu^*$ constitutes a mean-field equilibrium of the technology adoption game if $\mu^* \in \mc{BR}(\alpha(\mu^*))$.
\end{definition}

Note that a mean-field equilibrium is a tractable variation on our original finite network game.  In particular, agents are not required to hold complex beliefs about the adoption behavior of their neighbors, and the strategic description is relatively simple.
This allows us to obtain many useful results about the structure of equilibria.  We begin with a couple of background results stated
in the following theorem.

\begin{theorem}[Existence]
\label{th:existence}
There exists a mean-field equilibrium to the technology adoption game.
Moreover, if $\mu^*$ and $\mu'^*$ are mean-field equilibria
then $\alpha(\mu^*)=\alpha(\mu'^*)$.
\end{theorem}


\subsection{Characterizing Mean-Field Equilibria}
\label{sec:CharactMFE}

We now characterize the mean-field equilibria.    To state our main results we require the following definitions.

\begin{definition}[Double-threshold strategy]
\label{def:doublethreshold}
A mean-field strategy $\mu$ is a {\em double-threshold strategy} if there exist $d_L, d_U \in \mathbbm{N} \bigcup \{ \infty \}$,  such that:
\begin{align*}
d < d_L &\implies \mu(d) = 1;\\
d_L < d < d_U &\implies \mu(d) = 0;\\
d > d_U &\implies \mu(d) = 1.
\end{align*}
We refer to $d_L$ as the {\em lower threshold} and $d_U$ as the {\em upper threshold}.
\end{definition}
Note that if $d_L=0$ and/or $d_U = \infty$ then the strategy is effectively a {\em single} threshold
strategy or a constant strategy.
In a double-threshold strategy, agents adopt the technology early at the high and low degrees, and
between these cutoffs they free-ride by waiting to see what their neighbors learn.
The definition does not place any restriction on the strategy {\em at} the thresholds $d_L$ and $d_U$ themselves; as there may be indifference and randomization at these thresholds.

The following theorem establishes that every mean-field equilibrium involves an essentially unique double-threshold strategy.

\begin{theorem}[Double-threshold equilibrium]
\label{th:DoubleThreshEq}
If $\mu^*$ is a mean-field equilibrium then (i) it is a double-threshold strategy, and
(ii) the upper and lower thresholds are essentially unique: there exists a single pair $(d_L^*,d_U^*)$ that are valid thresholds for every  mean-field equilibrium.
\end{theorem}

Note that the same double-threshold strategy can have multiple representations; for example, if $\mu$ is a double-threshold strategy with $\mu(1) = 1$, $\mu(2) = 1$, and $\mu(3) = 0$, then the lower threshold can be either $d_L = 2$ or $d_L = 3$.  The theorem asserts that there exist a single pair of lower and upper thresholds that are valid for {\em every} possible mean-field equilibrium.

Note that in a double-threshold equilibrium, agents of high and low degree adopt early for different reasons. Low-degree agents adopt early because the benefits of informational free-riding are not high enough to justify the expected consumption lost by delaying adoption. On the other hand, high-degree agents adopt early because the expected referral rewards are high enough to overcome the benefits of informational free-riding.

The reasoning behind the mean-field equilibrium is seen quite easily from the utility function.
The difference in expected utility for an agent of degree $d$ of adopting early versus late is:
\begin{eqnarray}
\label{DeltaPi}
\Delta \Pi(\alpha,d) &=& \Pi^{adopt}(\alpha,d) - \Pi^{defer}(\alpha,d)  \\
  &=& p(A_0^H+A_1^H) + (1-p)A_0^L + \eta p (1- \alpha) d
 - p(1 - (1-\alpha)^d)A_1^H.\nonumber
\end{eqnarray}

We see the two competing terms of the benefit from adopting early --  the referral benefits that increase in $d$ linearly: $\eta p (1- \alpha) d$, and the
lost benefit of free riding which decreases in $d$ convexly: $  - p(1 - (1-\alpha)^d)A_1^H$.
These cross then (at most) twice, leading to the double threshold.

The following proposition establishes that there are conditions under which the mean-field equilibria involve just a single threshold.

\begin{proposition}
\label{cor:LowerThreshEq}
Let there be a maximal degree $\overline{d}>1$ such that $f(d)>0$ if and
only if $0<d\leq \overline{d}$.
There exist positive $\underline{\eta}$ and $\overline{\eta}$ such that
\begin{enumerate}
\item
if $\eta <\underline{\eta}$ then every mean-field equilibrium
can be characterized by $d^*_U> \overline{d}$,
\item
if $\eta >\overline{\eta}$
then every mean-field equilibrium can be characterized by $d^*_L=0$, and
\item
if $\underline{\eta}< \eta < \overline{\eta}$ then each mean-field equilibrium can be characterized by some $d^*_L\geq 1$ and $d^*_U\leq \overline{d}$.
\end{enumerate}
\end{proposition}

The equilibria are not completely ordered via $\eta$ since there are interactions between the upper and lower thresholds.
For instance, as more low degree agents adopt in the first period, it becomes less attractive for the high degree agents to adopt early.

Thus, Proposition \ref{cor:LowerThreshEq} states that for low enough referral incentive $\eta$, the equilibrium is effectively a lower-threshold strategy under which only lower-degree agents adopt early. For high enough referral incentive $\eta$, the equilibrium is effectively an upper-threshold strategy under which only higher-degree agents adopt early. For intermediate values of $\eta$, then the equilibrium is a double-threshold strategy with early adopters being both of low and high degree.

\subsection{Comparative Statics}
\label{sec:CompStat}

The model allows for comparative statics in the edge-perspective degree distribution $\tilde{f}(d)$. As we recall, this is simply the distribution over a neighbor's degree. The following propositions show that, depending on the value of $\eta$, a first-order stochastic dominance shift in the edge-perspective degree distribution can have opposite effects on the resulting expected adoption in equilibrium.

Let $\underline{\eta}(\tilde{f})$ and $\overline{\eta}(\tilde{f})$ denote the $\underline{\eta}$ and $\overline{\eta}$ such that Proposition \ref{cor:LowerThreshEq} holds, noting the dependence on the distribution $\tilde{f}$.

\begin{proposition}
\label{pr:FOSD_f_tilde}
Let  $\mu^*$ be any mean-field equilibrium under the edge-perspective degree distribution $\tilde{f}$, and {let $f' \neq f$ be a degree distribution such that the corresponding edge-perspective degree distribution $\tilde{f}'$ first order stochastically dominates $\tilde{f}$.  Further, for each degree distribution $f$ let $\underline{\eta}(\tilde{f})$ and $\bar{\eta}(\tilde{f})$ be the thresholds guaranteed by Proposition \ref{cor:LowerThreshEq}.}
\begin{itemize}
\item If $\eta <\underline{\eta}(\tilde{f})$ then any mean-field equilibrium $\mu'^*$ relative to $\tilde{f}'$ satisfies  $\alpha(\mu'^*) \leq \alpha(\mu^*)$.
\item If $\eta > \overline{\eta}(\tilde{f})$ then any mean-field equilibrium $\mu'^*$ relative to $\tilde{f}'$ satisfies $\alpha(\mu'^*) \geq \alpha(\mu^*)$.
\end{itemize}
\end{proposition}

The first part of the above proposition states that
if the equilibrium $\mu^*$ is a lower-threshold strategy
under $\tilde{f}$, then a first order stochastic dominance shift in $\tilde{f}$
leads to a new lower-threshold equilibrium $\mu'^*$ that has less informational
access $\alpha(\mu'^*)$. The intuition is that the new $\tilde{f}'$ puts
more weight on neighbors who delay adoption, thus lowering informational access.
The second part states that when the equilibrium $\mu^*$ is an upper-threshold strategy
under $\tilde{f}$, then a first order stochastic dominance shift in $\tilde{f}$ leads
to a new upper-threshold equilibrium $\mu'^*$ with more informational access $\alpha(\mu'^*)$.
The intuition is now that the new $\tilde{f}'$ puts more weight on neighbors who adopt early,
thus increasing informational access.

For intermediate levels of $\eta$, the equilibria cannot be ordered as
we shift the distribution, because under the double-threshold strategies weight
could be shifting to degrees that are either adopting or not as the distribution changes.

\section{Dynamic Pricing and Information Diffusion}
\label{SecDynamicPricing}

When the quality of the technology is uncertain, a monopolist wishing to market
the technology faces the risk of non-adoption by uninformed agents.
In other words, if an agent delays adoption in the hope of gathering information
and fails to obtain that information, he may elect not to adopt the technology at a
later stage. A dynamic pricing policy that properly encourages some agents to adopt
early can counteract this effect: because early adopters can spread information about the
quality of the technology, they decrease the fraction of agents who remain uninformed in the second period,
and consequently yield a higher adoption rate (if the quality of the good is high).
If the cost of inducing some
consumers to adopt early is low enough, this can increase the monopolist's profits.

We study an environment in which the monopolist does {not} know the topology of the social network linking consumers, and instead knows the distribution from 
which it is drawn.  Moreover, the monopolist knows the product quality $\theta$, whereas consumers do not (i.e., they simply have a prior $p$ about it). Thus, the actions of the monopolist may signal information about $\theta$.
However, before studying the full equilibrium, we first analyze the profits as a function of pricing strategy holding prior beliefs of the consumers fixed.  We then return to a full equilibrium analysis in Section \ref{sec:profitMax}.

\subsection{Dynamic Pricing Policies}
\label{Sec:Dynamic_Pricing_Policies}

To study pricing we need the following definition.

\begin{definition}
A dynamic pricing policy is a triplet $(P_0,P_1,\eta)\in \mathbbm{R} \times \mathbbm{R}_+ \times \mathbbm{R}_+$ consisting of prices for the first period, second period, and a referral fee.
\end{definition}

Note that we allow the first period price to be negative:
this allows the firm to pay early adopters - much as a company might pay a
high-degree celebrity to adopt its product, knowing that the individual is observed
by many others.   However, prices in the second period only make sense if they are
nonnegative, as the firm has nothing to gain by paying consumers in the second
period (this behavior does not provide any incentive to adopt earlier and does not help the second-period profit).
Lastly, we only allow nonnegative referral payments:  the firm
cannot charge consumers because others also adopt the product
(and this would not help with incentives or profits).\footnote{Our referral fee is not dependent upon a consumer's degree.  Allowing for nonlinear referral fees is explored in
independent work by \cite{lobel}, but without the learning effects or other pricing tradeoffs that are the focus of our paper.}

The payoffs of a consumer $i$ are:
\\
\begin{table}[h]
\begin{center}
  \begin{tabular}{ |l | c | r| }
  \hline
       & $\theta = H$ & $\theta = L \ \ \ \ \ \ \ \ \  $ \\ \hline
    \textrm{adopt at}\ t = 0 & $A_0^H + A_1^H + \eta X_{i,1} - P_0$ & $A_0^L  + \eta X_{i,1} - P_0 \ \  $ \\ \hline
    \textrm{adopt at}\ t = 1 & $A_1^H  - P_1$ & $A_1^L  - P_1 \ \ \  \ $ \\
    \hline
  \end{tabular}
  \caption[Payoffs under a Dynamic Pricing Policy]{Payoffs under a Dynamic Pricing Policy}
  \end{center}
  \label{tbl:payoffs_prices}
  \end{table}

We remark that early adopters only pay once: they pay $P_0$, but don't face another payment in the second period.  

We also remark that this set of pricing policies precludes offering a refund in the second period if the good turns out to be of low quality.  If a monopolist can
commit to refunds, then a perfect pricing policy becomes possible:
set $P_0=A_0^H+A_1^H$ and then offer a refund of $A_0^H-A_0^L+A_1^H$ if the
good turns out to be of low quality, and charge a price of $P_1\geq A_1^H$.
In that case, all consumers are willing to buy in the first period.  The monopolist earns exactly the full social surplus and the consumers pay their full value.
On the other hand, we contend that such a policy is not reasonable to consider.
The main issue is that commitment to refunds is not credible in general; if the
good turns out to be of low quality, then in the extreme the firm will simply
close and fail to offer refunds.  Thus, the pricing policies that we consider are the relevant ones.

Again, consumers' decisions are based on the differences in expected utilities from early versus late adoption.
Holding the prior $p$ fixed\footnote{In Section \ref{sec:profitMax} we show that
the prior $p$ is consistent with a pooling equilibrium, when pricing policies can be
used as signals of the product quality.}, this is
an easy variation on (\ref{DeltaPi}) in which we now add the different prices at different dates:

\begin{eqnarray}
\label{DeltaPi2}
\Delta \Pi(\alpha,d) &=& \Pi^{adopt}(\alpha,d) - \Pi^{defer}(\alpha,d)  \\
  &=& p(A_0^H+A_1^H) + (1-p)A_0^L - P_0 + \eta p (1- \alpha) d
 - p(1 - (1-\alpha)^d)(A_1^H - P_1) .\nonumber
\end{eqnarray}

Here note something that makes the pricing policy easy to understand.
The only term from the monopolist's policy that truly affects consumers
differently is the referral incentive benefit,  $\eta p (1- \alpha) d$, which is
increasing in degree.  The prices impact all consumers in the same way.
This feature provides a sort of ``crossing condition'' of preferences in referral incentives,
which is what enables it to be a screening device.

 Each pricing policy $(P_0, P_1, \eta)$ defines a particular dynamic game among
 the consumers and belongs
 to the same class as the one presented in Section \ref{SecModel}; i.e.,
 with a double-threshold equilibrium characterization. Thus, equilibria exist by a similar argument.
 This simple transformation of the original game then leads to a rich setup in which we can study the effect of various pricing policies on early adoption, information diffusion and free-riding.

Before examining the equilibrium under a profit-maximizing pricing policy, we state a couple of results about the nature of the equilibrium under certain types of pricing policies. The following proposition has important implications. It states that a dynamic pricing policy without referral incentives constitutes a mechanism under which only lower-degree agents choose to adopt early.

\begin{proposition}[Screening without referrals]
\label{obs:LowerThresholdMech}
Under a dynamic pricing policy $(P_0,P_1,0)$ (i.e., where $\eta = 0$), a mean-field equilibrium $\mu^*$ has a single threshold $d_L^*$ such that $\mu^*(d) = 1$, for $d < d_L^*$, and $\mu^*(d) = 0$, for $d > d_L^*$. In other words, such a policy constitutes a screening mechanism under which lower-degree agents adopt early while higher-degree agents free-ride on the information generated by the former.
\end{proposition}

A monopolist can thus guarantee that agents with degrees below a certain threshold will adopt early. The next proposition shows that for high enough $\eta$, a pricing policy with referral incentives constitutes a mechanism under which only higher-degree agents choose to adopt early.

\begin{proposition}[Screening with referrals]
\label{obs:UpperThresholdMech}
For any $P_0$, $P_1$, there exists $\eta^+ < \infty$ such that under a dynamic pricing policy $(P_0,P_1,\eta)$ where $\eta > \eta^+$, a mean-field equilibrium $\mu^*$ is such that $\mu^*(d) = 0$, for $d < d_U^*$, and $\mu^*(d) = 1$, for $d > d_U^*$. In other words, such a policy constitutes a screening mechanism under which higher-degree agents adopt early while lower-degree agents free-ride on the information generated by the former.
\end{proposition}

With extremely high referral rewards (high $\eta$), all agents would like to earn those
rewards,
but that cannot be an equilibrium as there would be no agents left to earn referrals from.
It is then easy to see that when the referral reward amount $\eta$ becomes high enough,
the unique equilibrium
is such that the lowest-degree agents
(e.g., degree 1) mix until they are indifferent, and all higher-degree agents
adopt early. A monopolist can thus guarantee that agents with degrees above a
certain threshold adopt early simply by pushing $\eta$ sufficiently high.

It is worth noting that the only network information
that the monopolist needs to implement the screening mechanisms described in
Propositions \ref{obs:LowerThresholdMech} and \ref{obs:UpperThresholdMech} is
basic information about the degree distribution\footnote{\cite{jimenezVersioning} also builds a model that relates seller decisions to the degree distribution, but in a different context.} $f(d)$. Our analysis thus
differs from papers like \cite{Candogan12} and \cite{Bloch13}, where a monopolist
incentivizes certain agents based on the full knowledge of the network topology -
and thus
 price discriminates based on network position rather than via screening as in our
 analysis.  Clearly, having more knowledge can increase the monopolist's profits,\footnote{
 How a monopolist's price adjusts with the precise network position of agents, in a setting
 with consumption complementarities across agents, is the subject of \cite{fainmesser}.
}
 but may not be available for many consumer goods.

\subsection{Information Diffusion}

We now introduce a few more useful definitions.

\begin{definition}[$\beta$-strategy]
Define $\beta: \mc{M} \rightarrow [0,1]$, by
\begin{equation*}
\beta(\mu) = \sum_{d \geq 1}  f(d) \mu(d)
\end{equation*}
A $\beta$-strategy $\mu \in \mathcal{M}$ is a mean-field strategy for which $\beta(\mu)=\beta$. In other words, it is a strategy that leads a fraction $\beta \in [0,1]$ of agents to adopt early (at time $t=0$). The set of $\beta$-strategies is denoted by $\mathcal{M}(\beta)$.
\end{definition}

The difference between $\beta(\mu)$ and $\alpha(\mu)$ is that the former is the expected fraction of early adopters (overall, from the monopolist's perspective), while the latter is the expected fraction of early adopters in a given consumer's neighborhood (so from the consumer's perspective, and thus weighted by the neighbor degree as a consumer is more likely to have higher-degree neighbors than from a random pick in the population).

Such a definition is useful when examining the informational access $\alpha(\mu)$ of a strategy. Indeed it allows us to compare it with other strategies under which the same fraction of agents adopt early and thus diffuse information about the quality of the technology. A given level of information achieved by a strategy under which a smaller fraction of agents adopt early can reasonably be thought of as more efficient, in an informational sense. This leads to the concept of informational efficiency, which is defined next.

\begin{definition}[Informational efficiency]
\label{def:InfoEfficiency}
The informational efficiency of a strategy $\mu \in \mathcal{M}$ is a mapping $\mathcal{E}: \mathcal{M} \rightarrow \mathbbm{R}^+$, which normalizes the informational access by the mass of agents generating information signals. It is expressed as

\begin{eqnarray}
\mathcal{E}(\mu) &=& \frac{\alpha(\mu)}{\beta(\mu)} \nonumber
\end{eqnarray}
\end{definition}

As one example of why informational efficiency is a useful metric, recall that first-period consumption has a negative
expected value. Thus, adopting in the first period provides an information gain from experimentation to the rest of the population.  Notably, it is beneficial to have the highest-degree individuals do that experimentation: this provides
a high informational access for a low fraction of experimenters, i.e., high informational efficiency.

It is clear from Proposition \ref{obs:LowerThresholdMech} that a dynamic pricing policy without referral incentives  leads to an adoption strategy $\mu^*$ with minimal informational efficiency (out of all strategies leading to a fraction $\beta(\mu^*)$ of early adopters). Indeed, it allocates the mass $\beta(\mu^*)$ of early adopters in a way that diffuses information in the worst possible manner:  it is the lowest-degree agents who adopt early. Such policies include the often-used price discounts given to early adopters.


In contrast, the informational efficiency achieved with a dynamic pricing policy involving referral incentives can be high. Indeed, from Proposition \ref{obs:UpperThresholdMech}, it is clear that a dynamic pricing policy with sufficiently high referral incentives allocates the mass $\beta^*$ of early adopters in a way that maximizes the diffusion of information, as it is the highest degree individuals who adopt early.


These observations have clear implications for a monopolist marketing the technology with a dynamic pricing policy. In fact, there is a cost associated with $\beta(\mu^*)$, the mass of early adopters, since those agents must be given an incentive to adopt early. The informational efficiency of the resulting strategy can therefore have an important effect on the profit that can be achieved. This is examined in the next section.


\section{Profit Maximization}
\label{sec:profitMax}

We now return to study the monopolist's pricing strategy and the agents' prior beliefs.

\subsection{Preliminaries: Equilibrium and Profit Definitions}

We
study an asymmetric information setting in which the monopolist is informed of the quality of the
technology (i.e., she observes $\theta$), but consumers are not (i.e., they do not observe $\theta$).

Pricing policies can thus be used as signals of the product quality: upon observing $(P_0,P_1,\eta)$,
consumers  now update their belief from the prior $p$ to a posterior belief $\mathbbm{P}\{\theta = H|P_0,P_1,\eta\}$ that reflects any additional information they may infer from the pricing policy used by the monopolist. We can thus replace $p$ by $\mathbbm{P}\{\theta = H|P_0,P_1,\eta\}$ in a consumer's expected payoff function (in (\ref{DeltaPi2})).
We examine cases in which consumers hold the same beliefs.

Consumers then play a mean-field equilibrium strategy $\mu^*$ in the dynamic adoption game that is induced, reflecting how the pricing policy $(P_0, P_1, \eta)$ and their updated belief affect their expected payoff.
We denote the set of such equilibria (as defined in the previous sections) by $\mc{EQ}(P_0,P_1,\eta)$, where the dependence on the pricing policy is made explicit.

We can now define a new dynamic game in the following way: (1) Nature chooses $\theta \in \{H,L\}$; (2) the monopolist then observes $\theta$ and chooses a pricing policy $(P_0,P_1,\eta)_{\theta}$; (3) the consumers then observe this pricing policy $(P_0,P_1,\eta)$, update their belief about the quality $\theta$ to a posterior belief $\mathbbm{P}\{\theta = H|P_0,P_1,\eta\}$ and then play an equilibrium in the ensuing dynamic adoption game.

We study a variation on the definition of perfect Bayesian equilibrium (e.g., \cite{watson2016}) adapted to our setting.
The variation is that we look at equilibria in which all consumers hold the same beliefs after seeing the pricing strategy of the monopolist, and then play a mean-field equilibrium in the resulting dynamic adoption game given their beliefs.  We refer to the resulting solution concept as a {\sl perfect Bayesian equilibrium}$^*$ (PBE$^*$).  A PBE$^*$ consists of a specification of a strategy for the monopolist, beliefs of the consumers for each pricing strategy, and a mean-field equilibrium in the dynamic adoption game given the pricing strategy and consumers' beliefs for each pricing strategy, such that:
\begin{enumerate}
\item the consumers' beliefs are obtained by Bayes' rule on the support of pricing strategies; and
\item the monopolist's strategy is a best response given the mean-field equilibrium
played as a function of the pricing strategy.
\end{enumerate}

Of course, determining the best response of the monopolist requires the latter to maximize profit,
which we now define precisely.  The profit of the monopolist is a function
$\pi: \mathbbm{R}^3  \times \{H,L\} \rightarrow \mathbbm{R}$, dependent on the
dynamic pricing policy $(P_0,P_1,\eta)$ and the state of the world $\theta$.
(To simplify notation we presume a marginal cost of zero for the product, but the analysis
extends easily to include a positive marginal cost.)

In defining the monopolist's profit, a technical challenge arises, which is seen as follows.
Suppose, for instance, that the monopolist set $P_0=\bar{A}$ and $P_1=A_1^H$ and $\eta=0$.
In that case, all consumers know they will get zero utility from any strategy that they follow and so
are completely indifferent between all of their strategies, regardless of their degree.  Which consumers should adopt early?  Which consumers should wait and free-ride?  Which should never adopt?   None of these questions are answered since all strategies lead to the same payoff.  If we allow the monopolist to choose the most favorable equilibrium, then she can ensure that she extracts the maximum possible surplus.  However, this is unlikely to emerge as the equilibrium in practice, since it relies on the ``knife edge'' of perfect indifference among consumers.

This technical issue is an artifact of modeling a pricing problem with a continuum of
prices --- which can leave all consumers exactly indifferent.
Charging prices on a discrete grid instead would (generically) tie down the
equilibrium (up to mixing by at most two types at the thresholds).
Thus, an appropriate way to solve this issue is to consider equilibria at the limit
of some sequence of equilibria that are robust to some
perturbations of the pricing grid.\footnote{See \cite{simonz1990,jacksonssz2002} for some discussion of this (which can also lead to existence issues in some settings, although not here).}
Here, it is enough simply to require that prices approach the limit prices from below, which guarantees that at least some types are not indifferent between all strategies at any sequence close enough to the limit.  This is the approach we use to define the profit of the monopolist.

We begin with a definition for the maximum profit among all equilibria at a fixed pricing policy:
\begin{equation}
\label{eq:fixed_policy_profit}
 \rho(P_0, P_1, \eta, \theta) = \sup_{\mu^* \in \mc{EQ}(P_0, P_1, \eta)} \left\{ \beta(\mu^*) P_0 + \gamma_{\theta}(\mu^*) P_1 - \phi_{\theta}(\mu^*) \eta \right\},
\end{equation}
where:
\begin{equation}
\label{eq:gammadef}
\gamma_H(\mu^*) = \sum_{d \geq 1} f(d)\cdot (1 - \mu^*(d) ) (1-(1-\alpha(\mu^*))^d) \text{ \ and \ } \gamma_L(\mu^*)=0
\end{equation}
is the equilibrium fraction of late adopters, when $\theta = H$ and $\theta = L$ respectively,  and
\begin{equation}
\phi_H(\mu^*) = (1-\alpha(\mu^*))\sum_{d \geq 1} f(d)  \mu^*(d) d \text{ \ and \ } \phi_L(\mu^*)=0
\end{equation}
is the expected number of referral payments that must be paid out to early adopters, when $\theta = H$ and $\theta = L$ respectively (recalling that an agent receives a referral payment for each late adopting neighbor).

Next we define the profit at a given pricing policy by taking the limsup of profit among pricing policies whose prices $P_0$ and $P_1$ approach the given policy from below, and whose referral incentive $\eta$ approaches it from above:
\begin{equation}
\pi(P_0,P_1,\eta,\theta) = \limsup_{P_0' \uparrow P_0, P_1'\uparrow P_1, \eta'\downarrow \eta  } \rho(P_0', P_1', \eta',\theta).
\label{eq:FullProfit}
\end{equation}

\subsection{Pooling and Separating Equilibria}

As there are two types for the product (low or high), there are two natural sorts of equilibria:
separating (in which each type of monopolist chooses a different pricing strategy) or pooling
(in which each type of monopolist chooses a common pricing strategy).\footnote{There also exist some semi-pooling equilibria that are trivial variations on the separating equilibria, but have equivalent profits as will be seen a bit later.}
In this section, we point out that all equilibria that involve separation are trivial,
while there are many pooling equilibria.  We argue that it is reasonable to focus attention on the
pooling equilibrium where the monopolist of the high type makes maximum profit.
First, this is obviously a natural benchmark to consider; the high quality seller is maximizing profits,
while the low quality seller makes as much as possible (namely, the same as the first
period profit of a high quality seller) -- and so this is the equilibrium that is most attractive for the sellers, who are choosing the pricing policies.
Second, by studying this benchmark, we are later able to obtain structural insight into
the dependence of the monopolist's pricing policy on model primitives (e.g., the degree distribution).
This pooling equilibrium is the object we study in the remainder of our analysis.

The following proposition is central to our analysis.

\begin{proposition}[Equilibrium]
\label{prop:poolingEq}
Let $(\hat{P}_0,\hat{P}_1,\hat{\eta})$ be a pricing policy such that $\pi(\hat{P}_0, \hat{P}_1, \hat{\eta}, H) \geq 0$ and $\pi(\hat{P}_0, \hat{P}_1, \hat{\eta}, L) \geq 0$; in other words, both the low and high type monopolists make nonnegative profit.

Then the monopolist choosing pricing policies $(P_0,P_1,\eta)_H = (P_0,P_1,\eta)_L =(\hat{P}_0,\hat{P}_1,\hat{\eta})$ in each state of the world $\theta \in \{H,L\}$ and consumers playing $\mu^* \in \mc{EQ}(\hat{P}_0,\hat{P}_1,\hat{\eta})$ with posterior belief $\mathbbm{P}\{\theta = H|\hat{P}_0,\hat{P}_1,\hat{\eta}\}=p$, and otherwise playing $\mu^*=0$ (not buying at all) with posterior belief $\mathbbm{P}\{\theta = H|P_0,P_1,\eta\}=0$ for any other policy $(P_0,P_1,\eta) \neq(\hat{P}_0,\hat{P}_1,\hat{\eta})$ constitute a pooling PBE$^*$.

On the other hand, there are no separating PBE$^*$ in which either type of the monopolist earns positive profits.
\end{proposition}

The nonexistence of nontrivial separating equilibria is easy to see.  The low type must earn 0 profits, since it is known to be low given the separation.
This means that the first-period strategy of the high type cannot earn positive profits, or the low type would mimic.  But any strategy that earns positive profits for the high type must have
a positive price in the first period as otherwise the consumers would all buy in the first period and not expect to pay anything for the product.  But this is then a contradiction.

Since there are no nontrivial separating equilibria, we are left only with separating equilibria that earn 0 profits for both types.  Such equilibria exist.  For instance, the high types can charge zero prices in both periods and offers no referral incentives, and any other strategy leads to beliefs of a low type.  The low type charges some other prices and referral incentives and gets no consumers.
Note that there are also some semi-pooling equilibria of this form, as the low type could also mix and sometimes mimic the high type, and the low type would earn 0 profits in either case.

The more interesting equilibria are the pooling equilibria: the monopolist chooses the same
strategy in either state of the world. Thus, agents cannot infer anything from the policy, so
their posterior belief is simply equal to their prior $\mathbbm{P}(H|P_0,P_1,\eta)=p$.  (This is convenient since we wish to highlight the effect of a dynamic pricing policy on information diffusion among agents rather than focusing on the signaling aspects of prices,
a subject that is already well studied (e.g., see \cite{riley2001}) and would obscure the analysis.)
The construction of the pooling equilibrium is quite intuitive.  If a monopolist with either type of product
behaved differently from the equilibrium strategy in either period,
the consumers would believe that the product is of low quality, so there would be
no purchases, which cannot be payoff-improving.  The beliefs of the consumers
are correct at all nodes reached with positive probability, and so the equilibrium conditions are satisfied.


Given the multiplicity of pooling equilibria, we focus on one that seems particularly natural\footnote{We do not see any reasonable refinement that picks any of the pooling equilibria compared to others.}, in which the High type maximizes profits and the low type mimics the high type to earn as much as possible subject to pooling and the information asymmetry.   The high-quality monopolist thus optimizes the choice of the policy parameters
knowing that the consumers' posterior belief will equal their prior $p$.  Not only is focusing on this equilibrium natural, it also allows us to carry out comparative statics as we vary aspects of the model (particularly the degree distribution).

Formally, we let
\begin{equation}
\label{eq:maxprofit}
\hat{\pi}= \max_{P_0, P_1, \eta} \ \pi(P_0,P_1,\eta,H)
\end{equation}
denote the maximal profit achievable over all pricing policies $(P_0,P_1,\eta)$ by a monopolist
with a high-quality product.

This optimal profit is difficult to compute in closed form as it
depends on the degree distribution in complex ways.  Effectively, the
monopolist would like to have the early adoption performed by high-degree
individuals. She can achieve that by charging prices that make consumers nearly
indifferent and by offering a referral incentive that is high enough.
The complicating factor is that some very low-degree
consumers might have such a low expectation of learning from a
neighbor that it is best to also entice them to consume in the first
period.  The potential optimal policies thus breaks down into a variety of
cases based on parameters and the degree distribution.  Therefore, in what
follows we focus on some variations on networks that provide
the basic intuitions; in addition, we focus our attention on when simplified policies --- those which use only a price discount, and those which use no discount but add a referral incentive --- can be
optimal.

\subsection{Comparing Price Discounting versus Referral Incentives}

We next discuss two restricted classes of dynamic pricing policies and compare them to the unrestricted
profits.  The policies that we consider are (i) two-price policies and (ii) referral incentives.
This allows us to see the tradeoffs between price discounts and referral incentives.

In the first class of policies, the monopolist uses inter-temporal price discrimination and charges different prices to early and late adopters. As will be seen later, a "discounted" price $P_0 \leq P_1$ is charged to early adopters in order to encourage them to adopt early and reveal information about the product to their neighbors.\footnote{Note that
 even though the early adopters will get an extra period of consumption, by assumption their expected utility in the first period is negative and the only value
  to the first period consumption is the learning value - by Assumption \ref{as:AlwaysFreeriding}.} Late adopters then pay the "regular" price $P_1$. In the second class of policies, a referral incentive $\eta$ is given to early adopters for each neighbor who adopts after them. All agents pay a "regular" price $P=P_0=P_1$, whether they adopt early or late. We will thus let $\mathcal{D} = \{(P_0,P_1,\eta): \eta = 0 \}$ denote the set of two-price policies and $\mathcal{R} = \{(P_0,P_1,\eta): P_0=P_1 \}$ denote the set of referral incentive policies.

We can thus restrict the profit function (\ref{eq:FullProfit}) to those two policy classes. The profit of a price discount policy (a $\mathcal{D}$-policy) can be written as

\begin{equation}
\label{eq:Profit_D}
\pi_{\mathcal{D}}(P_0,P_1,\theta) = \limsup_{P_0' \uparrow P_0, P_1'\uparrow P_1} \rho(P_0', P_1', 0,\theta),
\end{equation}

The profit of a referral incentive policy (a $\mathcal{R}$-policy), on the other hand, can be written as
\begin{equation}
\label{eq:Profit_R}
\pi_{\mathcal{R}}(P,\eta,\theta) = \limsup_{P' \uparrow P, \eta'\downarrow \eta} \rho(P', P', \eta',\theta)
\end{equation}

In what follows, we focus our attention on the monopolist's profit when $\theta=H$. The profit maximization problem can then be stated as

\begin{equation}
\label{ProfitMaxDiscount}
\begin{aligned}
& \underset{P_0,P_1}{\text{maximize}} \ \  \pi_{\mathcal{D}}(P_0,P_1,H) \\
 \end{aligned}
\end{equation}
in the first case and
\begin{equation}
\label{ProfitMaxReferrals}
\begin{aligned}
& \underset{P,\eta}{\text{maximize}} \ \  \pi_{\mathcal{R}}(P,\eta,H) \\
 \end{aligned}
\end{equation}
in the second case.

Since we focus our attention on $\mc{D}$- and $\mc{R}$-policies, analogous to \eqref{eq:maxprofit} we define maximum profit restricted to those policy classes:
\begin{equation}
\label{eq:maxprofit_TP}
\hat{\pi}_{\mathcal{D}} =\max_{P_0, P_1} \pi_{\mathcal{D}}(P_0,P_1,H)
\end{equation}
for two-price policies, and
\begin{equation}
\label{eq:maxprofit_ref}
\hat{\pi}_{\mathcal{R}} =\max_{P, \eta} \pi_{\mathcal{R}}(P,\eta,H)
\end{equation}
for referral policies.

In the next section, we study how these policies perform.  In particular, we study whether they are optimal over the \textit{whole} space of policy triplets $(P_0,P_1,\eta)$, both theoretically (for specific network structures) and numerically (more generally).

\subsection{Theoretical Results}

We start with a characterization of an optimal two-price policy for general degree distributions.

\begin{proposition}[Optimal two-price policy]
\label{prop:opt_two_price_pol}
For any degree distribution $f(d)$, $(\hat{P}_0,\hat{P}_1,0)=(\bar{A},A_1^H,0)$ is always an optimal two-price policy, i.e. $\hat{\pi}_{\mathcal{D}} = \pi_{\mathcal{D}}(\bar{A},A_1^H,H)$.
\end{proposition}

Thus, a monopolist can always achieve optimal profits by charging prices $\hat{P}_0$ and $\hat{P}_1$ that reap the adopters' full surplus in each period. The intuition is that any lower-threshold adoption strategy $\mu^*$ (i.e. any $d_L^*$) can be achieved by an appropriate choice of sequences $P_{0,k}$ and $P_{1,k}$ converging to $\bar{A}$ and $A^H_1$ from below. Hence, by the definition of profits in Eq. (\ref{eq:Profit_D}), it follows that the monopolist can charge such maximal prices and achieve the optimal lower-threshold adoption strategy leading to the optimal fractions of early and late adopters for some particular $f(d)$.

Note that the optimal referral policy $(\hat{P},\hat{P},\hat{\eta})$ is more difficult to characterize than the optimal two-price policy.  Indeed a referral policy can lead to a double-threshold adoption strategy, which may be optimal if there is a large mass of low-degree agents. While the referral $\hat{\eta}$ would mainly serve to attract high-degree agents (and set the ideal upper-threshold $d_U^*$), the monopolist could then charge a low price $P$ to entice low-degree agents to adopt early (since their low degree would likely prevent them from collecting information from their friends and thus they would not adopt late). If there is a large mass of high-degree agents, however, the monopolist may prefer to use referrals to achieve an upper-threshold strategy. In this case he could charge the highest possible price $P=A_1^H$ and use referrals to entice high-degree agents to adopt early. Therefore, there is no straightforward analogue to Proposition \ref{prop:opt_two_price_pol} for general degree distribution $f(d)$.

We will now examine the performance of those two classes of dynamic pricing policies $\mc{D}$ and $\mc{R}$ on $d$-regular networks, i.e., on networks in which all agents have degree $d$. This models a case in which agents are homogeneous in their propensity to interact with others. The following theorem states results for that particular case.

\begin{theorem}[Optimal profit on d-regular networks]
\label{th:discount_opt_d_reg}
Suppose the network is $d$-regular, i.e., $f(d) = 1$ for some $d$ and $f(d)=0$ otherwise. Then:
\begin{itemize}
\item[(i)] For all $d$, $\hat{\pi}_{\mathcal{D}} = \hat{\pi}  > \hat{\pi}_{\mathcal{R}}$;
\item[(ii)] $\lim_{d \to \infty} \hat{\pi}_{\mathcal{D}} = \lim_{d \to \infty}  \hat{\pi}_{\mathcal{R}}=\lim_{d \to \infty} \hat{\pi} = A_1^H$.
\end{itemize}
\end{theorem}


Part (i) states that on $d$-regular networks, an optimal two-price policy is optimal over the \textit{whole} space of policy triplets $(P_0,P_1,\eta)$. The intuition behind this result is that with a two-price policy, a monopolist can capture the full surplus of both early and late adopters while being also able to choose the optimal informational access $\alpha^*$.
\begin{figure*}
\centerline{
\includegraphics[scale=0.5]{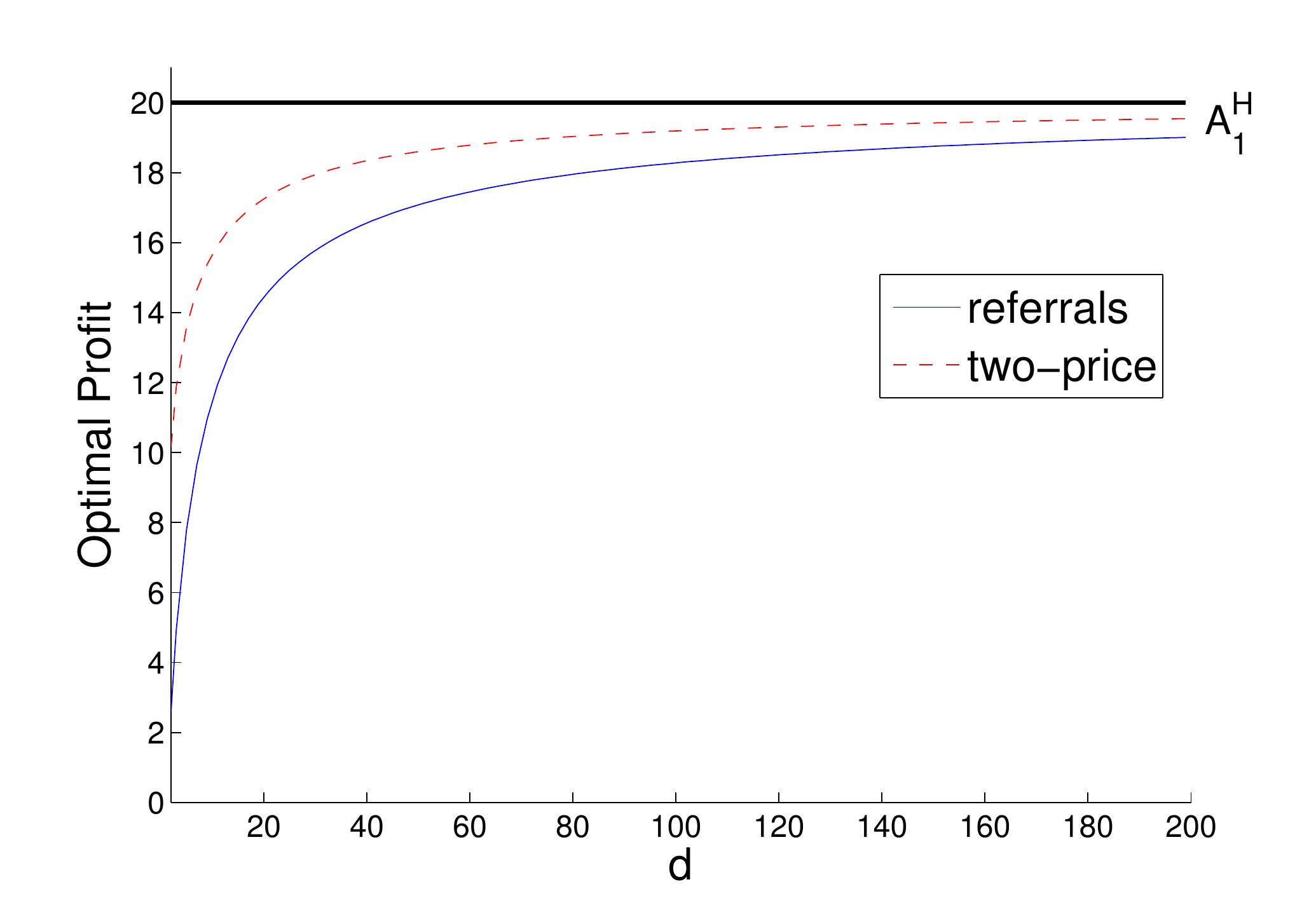}
}
\caption{Optimal Profit of Two-Price and Referral Incentive Policies on a $d$-regular network with degree $d$. \textit{Model parameters are $A_0^H=10$, $A_1^H=20$, $A_0^L=-10$, $A_1^L=-20$ and $p=0.4$. $A_1^H$ is shown by the horizontal black line.}}
\label{Profit_d_regular}
\end{figure*}
This is shown in Figure \ref{Profit_d_regular}, where the profits under optimal policies of classes $\mc{D}$ and $\mc{R}$ are plotted against the degree $d$. In particular, the profit achieved by any referral policy is strictly dominated by the profit achieved by a price discount-optimal policy. The reason is that whether a referral reward is paid out by the monopolist depends on the technology's quality $\theta$. If $\theta=L$, then an early adopter will {\em not} reap any referral rewards since none of his neighbors will adopt after him. A price discount $P_1-P_0$, on the other hand, is paid to an early adopter regardless of the value of $\theta$. A monopolist must thus offer an ``inflated" referral reward amount $\eta$ that compensates an early adopter for the risk of not receiving it. Since profit is optimized by the monopolist conditional on $\theta = H$, this results in referrals being a more costly form of incentives than inter-temporal price discrimination.

Part (ii) states that as the network becomes fully connected, both $\mc{D}$-optimal and $\mc{R}$-optimal policies become equivalent.  In particular, in this limit even referral policies are optimal over the space of all possible policies.  Essentially, in this case, the consumers expect to be nearly perfectly informed by the second period
even if there is just a tiny fraction of early adopters.  Thus, the equilibrium involves only a tiny fraction adopting in the first period (the optimal
policy makes consumers indifferent) -- then whether they are paid via discounts or referrals is a negligible difference -- and a full surplus of $A^H_1$ is extracted in the second period from late adopters. This tends to be \textit{all} of the agents as $d$ grows,
leading to full surplus and full efficiency.

Theorem \ref{th:discount_opt_d_reg} has important implications for a marketer: In an environment in which agents have (roughly) the same propensity to interact with each other, then a two-price policy is the optimal choice of policy. On the other hand, in a (roughly) fully-mixed matching environment, both two-price and referral incentive policies perform well (if suitably chosen).

\begin{figure*}
\centerline{
\includegraphics[scale=0.5]{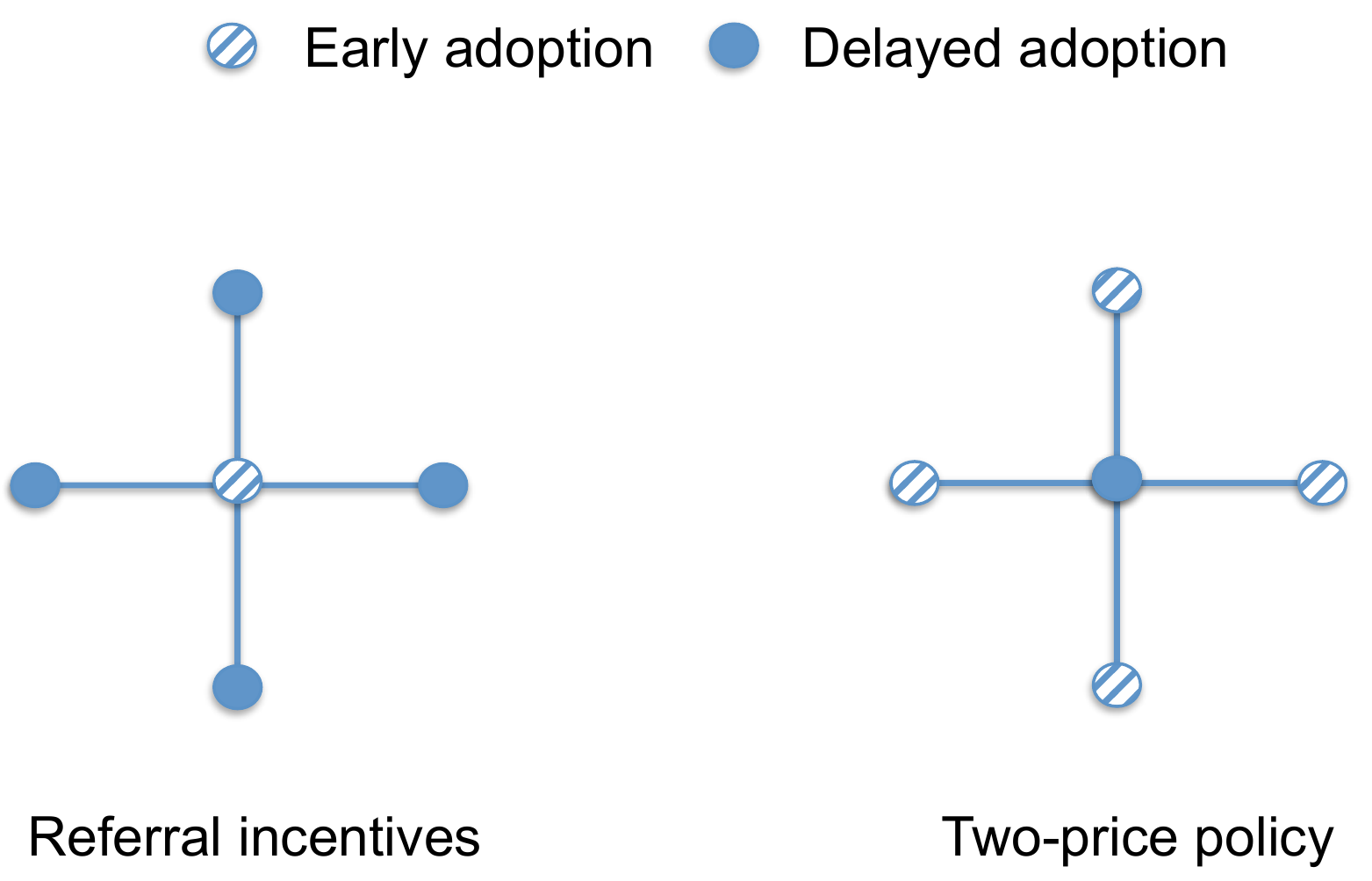}
}
\caption{Early Adoption Pattern under Referrals and Two-Price Policies on Star Networks.}
\label{fig:stars}
\end{figure*}

The preceding insights are obtained under the assumption of a regular network. To gain insight into the role of degree heterogeneity, we analyze a two-degree network, i.e., a network in which agents can either be of low degree $d_l$ or of high degree $d_u$.  We show that under ideal conditions, this heterogeneity allows a monopolist to devise a strategy of maximal informational access with minimal cost. To see the intuition, consider the simple example of a star network in Fig. \ref{fig:stars}.
We see that the different patterns of early adoption can have significant effects on information diffusion. The lower-threshold equilibrium in the case of a two-price policy implies that the price discount is given to many agents (all the periphery nodes) and achieves only limited informational diffusion (only the center node is informed and adopts late). This is thus costly for the monopolist as a small fraction of agents adopt late and a large fraction adopt early and must be incentivized. On the other hand, the upper-threshold equilibrium in the case of an appropriately-chosen referrals policy is more informationally efficient.
Which of these ends up being more profitable depends on the size of the price discount and the referral incentives.  In cases where there are some very high degree nodes
present in the network, however, it becomes quite cheap to offer the referral incentives, as we show in the following proposition.

\begin{proposition}[Optimal profit on two-degree networks]
\label{pr:2d_limits}
If the network has only two degrees, i.e., $f(d_u) = q$ and $f(d_l)=1-q$ for some $d_u \geq d_l$,
then, for any $d_l$:
\begin{enumerate}
\item[(i)]
\[ \lim_{q \to 0} \lim_{d_u \to \infty} \hat{\pi}_{\mathcal{D}} < \lim_{q \to 0} \lim_{ d_u \to \infty} \hat{\pi}_{\mathcal{R}} = \lim_{q \to 0} \lim_{d_u \to \infty} \hat{\pi}; \]
\item[(ii)]   $(\hat{P}_0,\hat{P}_1,0)= (\bar{A},A^H_1,0)$ and $(\hat{P},\hat{P},\hat{\eta}) = (A^H_1,A^H_1,\eta^+)$, where $\eta^+= \frac{ A_1^H - \bar{A}}{p(1-\tilde{f}(d_u))d_u}$, are optimal two-price and referral policies, respectively, as $q \rightarrow 0$ and $d_u \rightarrow \infty$;
\item[(iii)]
\[ \lim_{q \to 1} \lim_{d_u \to \infty} \hat{\pi}_{\mathcal{D}} = \lim_{q \to 1} \lim_{ d_u \to \infty} \hat{\pi}_{\mathcal{R}} = \lim_{q \to 1} \lim_{d_u \to \infty} \hat{\pi}; \]
\item[(iv)] For any $q \in (0,1)$, $\lim_{d_u \to \infty} \hat{\pi}_{\mathcal{R}} < A_1^H.$
\end{enumerate}
\end{proposition}

The preceding proposition examines \mc{D}- and \mc{R}-optimal profits as the higher degree $d_u$ grows arbitrarily large while the lower degree $d_l$ remains fixed.  In this case a small fraction of agents have a disproportionately large propensity to interact with others.  Part (i) states that referral policies are optimal over the whole space of policy triplets $(P_0, P_1, \eta)$, and in particular dominate discount policies. Note that a star network is one possible realization under this degree distribution.
To see this, consider the extreme case where $d_l=1$; informally, this corresponds to a star network with a single infinite-degree node at the center and an infinite number of degree-$1$ nodes in the periphery. A monopolist would thus want to incentivize {\em only} the agent at the center of this network. A referral incentive policy allows him to do that and thus to achieve maximum informational access ``for free": the total incentivizing cost can be shown to converge to zero while the total revenue can be shown to converge to $A^H_1$, the total surplus of late adopters and the maximum profit achievable.  On the other hand, the optimal two-price policy leads to a non-trivial fraction of degree-$1$ agents adopting early (because of the lower-threshold adoption strategy), a significant loss to the monopolist.

Part (ii) shows the shape of optimal limiting policies. The optimal two-price policy, as usual, sets prices reaping the full surpluses of early and late adopters so as to attain the optimal fraction of early adopters. Here this means \textit{all} low-degree agents adopting early and thus only \textit{their} surplus being reaped. Indeed, since a lower-threshold strategy necessarily leads to zero informational access in the limit, any low-degree agent delaying adoption would remain uninformed and thus not adopt late. The monopolist thus only captures the full surplus of early adopters, i.e. $\bar{A}$. The optimal referral policy, on the other hand, sets $\eta$ so as to make all high-degree agents adopt early, thus generating full informational access $\alpha^*=1$ with a vanishing fraction of early adopters. This allows her to reap the full surplus of late adopters $A_1^H$, all of whom are low-degree agents.

Part (iii) states that as the degree distribution converges to that of a fully connected network, profits under both $\mc{D}$- and $\mc{R}$-optimal policies are equal to the optimal profit over the set of all policies $(P_0,P_1,\eta)$. Indeed, in this case, we recover the result from part (ii) of Theorem \ref{th:discount_opt_d_reg}.  On the other hand, we see in part (iv) that as long as the fraction of higher-degree agents is non-trivial, a referral policy cannot capture the total surplus of late adopters because a non-trivial fraction of agents must be incentivized to adopt early.

\subsection{Numerical Investigation}

\noindent \textbf{More General Degree Distributions}

\begin{figure*}
\centerline{
\includegraphics[scale=0.5]{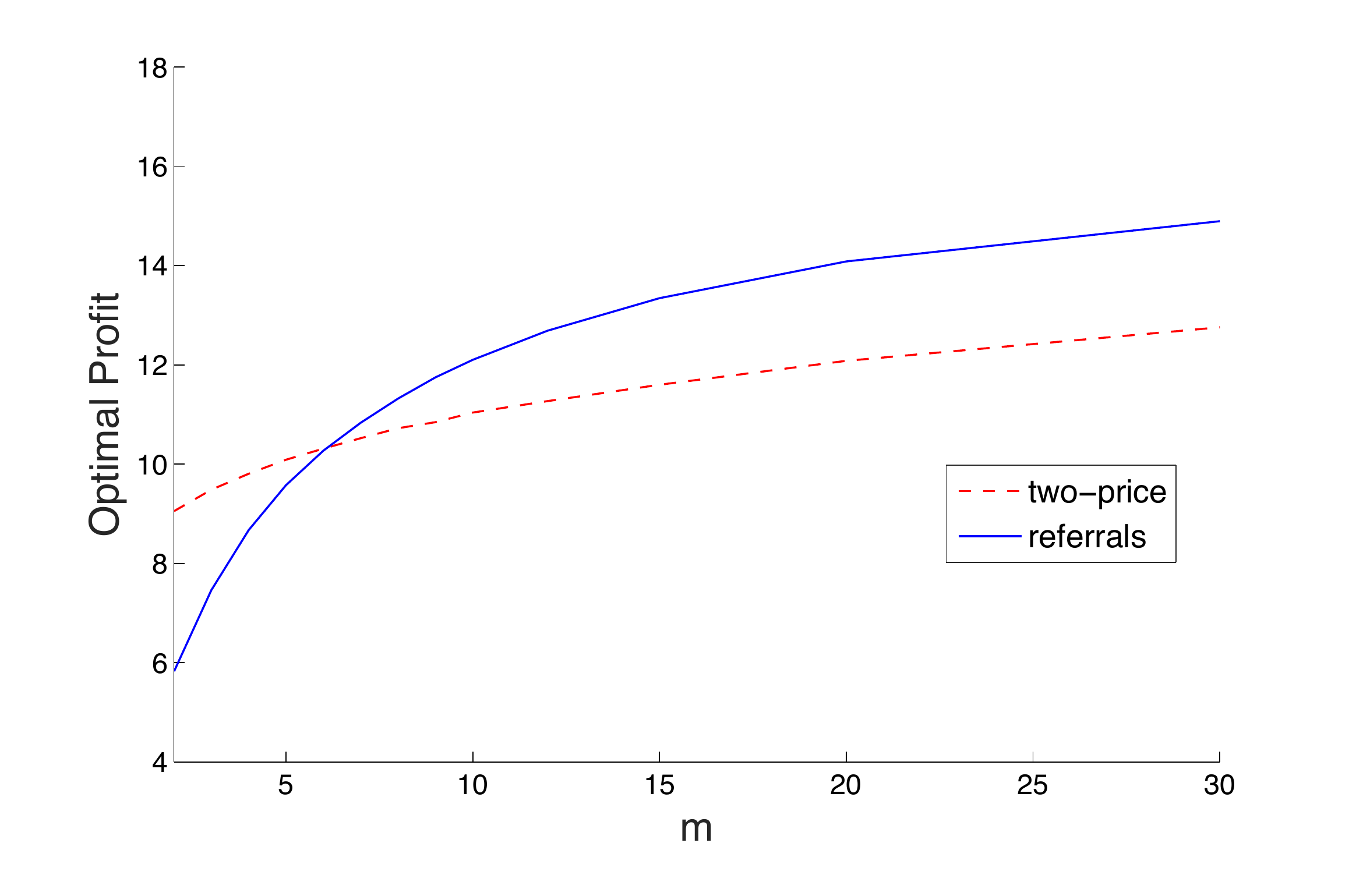}
}
\caption{Optimal Profit of Two-Price and Referral Incentive Policies vs Average Degree $m$. \textit{The degree distributions are as in (\ref{eq:JacksonRogersDistribution}) with $r=2$ in all cases. Model parameters are $A_0^H=10$, $A_1^H=20$, $A_0^L=-10$, $A_1^L=-20$ and $p=0.4$.}}
\label{ProfitPlotRefDisc}
\end{figure*}

To see how our theoretical results extend to more general degree distributions, we examine the performance of the two classes of dynamic pricing policies \mc{D} and \mc{R} for a family of degree distributions.  We use a model of \cite{JacksonRogers07}, which fits well across a wide range of social networks, and covers both scale-free networks and networks formed uniformly at random as extreme cases. The cumulative distribution function is
\begin{equation}
\label{eq:JacksonRogersDistribution}
F(d) = 1 - \Big(\frac{rm}{d+rm} \Big)^{1+r}
\end{equation}
where $m$ is the average degree and $0<r<\infty$. The distribution approaches a scale-free (resp., exponential) distribution as $r$ tends to $0$ (resp., $\infty$). This family has two interesting properties: varying $m$ is equivalent to a first-order stochastic dominance shift in the distribution; and varying $r$ is equivalent to a second-order stochastic dominance shift in the distribution.  Formally, when distribution $F'$ has parameters $(m',r')$ and distribution $F$ has parameters $(m,r)$ such that $r'=r$ and $m' > m$, then $F'$ strictly first-order stochastically dominates $F$.  Further, when distribution $F'$ has parameters $(m',r')$ and distribution $F$ has parameters $(m,r)$ such that $m'=m>0$ and $r' < r$, then $F'$ is a strict mean-preserving spread of $F$.
Fig. \ref{ProfitPlotRefDisc} illustrates the performance of \mc{D}- and \mc{R}-optimal pricing policies.
Each point on the curves represent the profit optimized\footnote{Note that, in this example, we only consider pricing policies under which at most one degree plays a mixed strategy, which is natural in this setting.} numerically as in (\ref{ProfitMaxDiscount}) and (\ref{ProfitMaxReferrals}), for the given degree distribution\footnote{Note that we consider a degree support from $1$ to $200$ for the degree distribution in (\ref{eq:JacksonRogersDistribution}). The latter is thus normalized over that finite support so that it sums to $1$.} (i.e., for each separate value of $m$, when $r$ is fixed to $2$). We see that referral incentive policies fare better than two-price policies on degree distributions with average degree $m$ \textit{higher} than a certain threshold ($m \approx 7$). Indeed on such distributions, the effect of Proposition \ref{obs:UpperThresholdMech} can be felt: agents with degrees above threshold $d_U^*$ invest early, and do so with maximum efficiency. They therefore generate high informational access at a low incentivizing cost (in the form of referrals payments) since the informational efficiency is maximal. Although the free-riding agents have degrees below a threshold $d_U^*$, they still have relatively high degrees since the distribution has a high $m$. They are thus very likely to collect information (i.e., have at least one neighbor who invested early). This translates into high rates of late adoption among free riders for a small fraction of highly-efficient early adopters.

A feature that emerges from Fig. \ref{ProfitPlotRefDisc} is that referral incentive policies are dominated by two-price policies on degree distributions with average degree $m$ \textit{lower} than a certain threshold ($m \approx 7$). On such distributions and under a referral incentive policy, free-riding agents have very low degrees. Thus, even if the early-adopting agents have higher degrees and thus generate high informational access, some free-riding agents are very likely to remain uninformed, because their low degrees translate into a low probability of having an early-adopting neighbor. This then translates into lower rates of late adoption. Under a two-price policy, however, it is agents with degrees below a threshold $d_L^*$ who adopt early. Although the latter tend to generate lower informational access, the free-riding agents have degrees above the threshold $d_L^*$ and are thus less likely to remain uninformed. This translates into an inefficient pattern of early adoption, but this is compensated by a reasonable rate of late adoption. In other words, a two-price policy dominates a referral incentive policy not because it generates higher informational access (and efficiency) but rather because it guarantees adoption by low-degree agents---precisely those agents who would likely fail to collect information from their neighbors under a referral incentive policy.

\begin{figure*}
\centerline{
\includegraphics[scale=0.8]{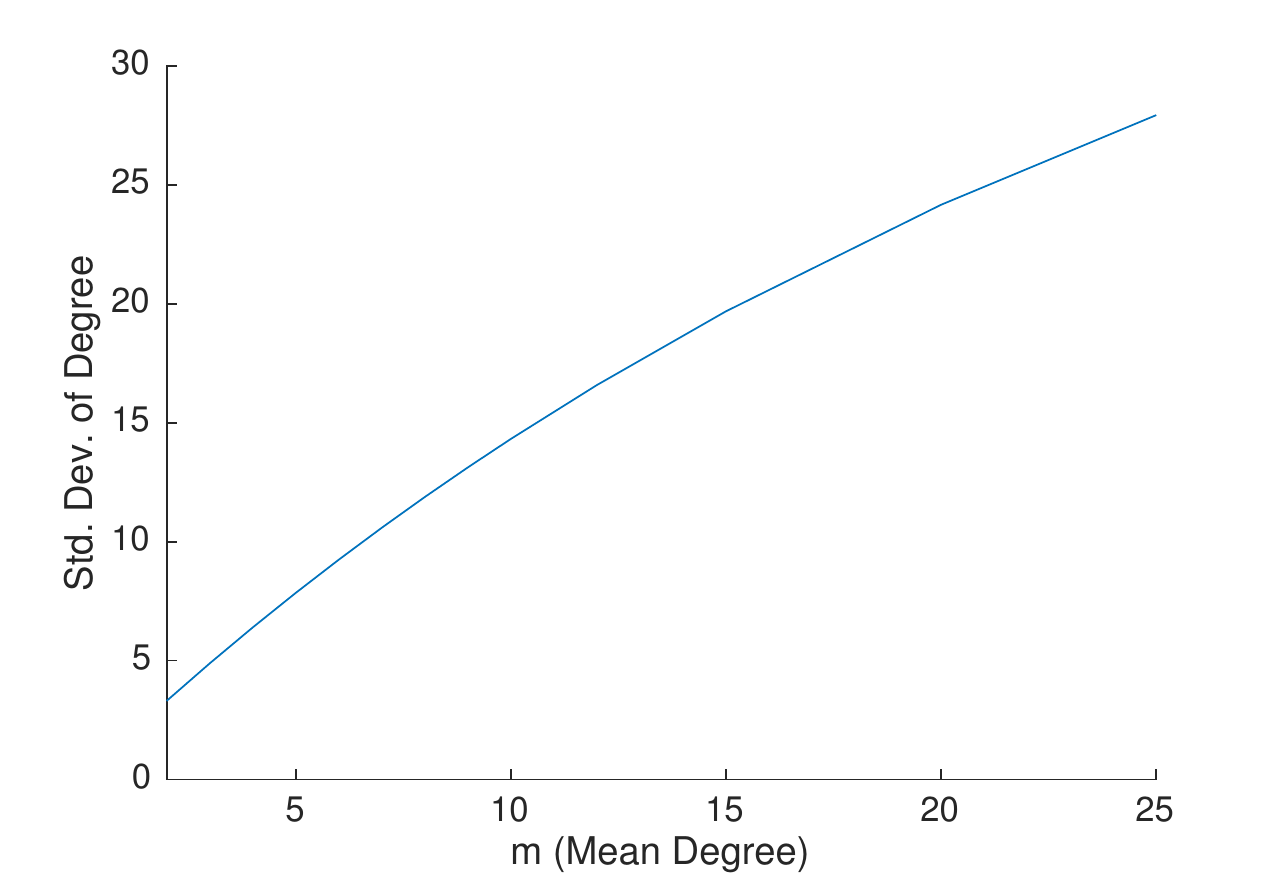}
}
\caption{Standard Deviation of Degree versus Average Degree $m$ for the Degree Distributions Used in Fig. \ref{ProfitPlotRefDisc}, i.e., as in (\ref{eq:JacksonRogersDistribution}) with $r=2$.}
\label{std_dev_versus_m}
\end{figure*}

It is important to note that the average degree $m$ itself does not drive the results seen in Fig. \ref{ProfitPlotRefDisc}. 
Indeed, we know from Theorem \ref{th:discount_opt_d_reg} that on a $d$-regular network, referrals do not dominate two-price policies, even when the average degree $d$ is increased. With this distribution, as $m$ increases so does the variance of the degree distribution.  High degree variance (i.e., high heterogeneity in degree) can allow referrals  to incentivize a small fraction of high-degree agents to adopt early and thus to spread information efficiently.   In Fig. \ref{std_dev_versus_m}, we plot the standard deviation of the degree versus the mean degree $m$ for the distributions used to produce the results of Fig. \ref{ProfitPlotRefDisc}. The standard deviation of the degree increases with $m$. This increased degree heterogeneity is a large part of what drives the profit results of Fig. \ref{ProfitPlotRefDisc}.

\

\noindent \textbf{Effect of the Variance of the Degree Distribution}

\begin{figure*}
\centerline{
\includegraphics[scale=0.8]{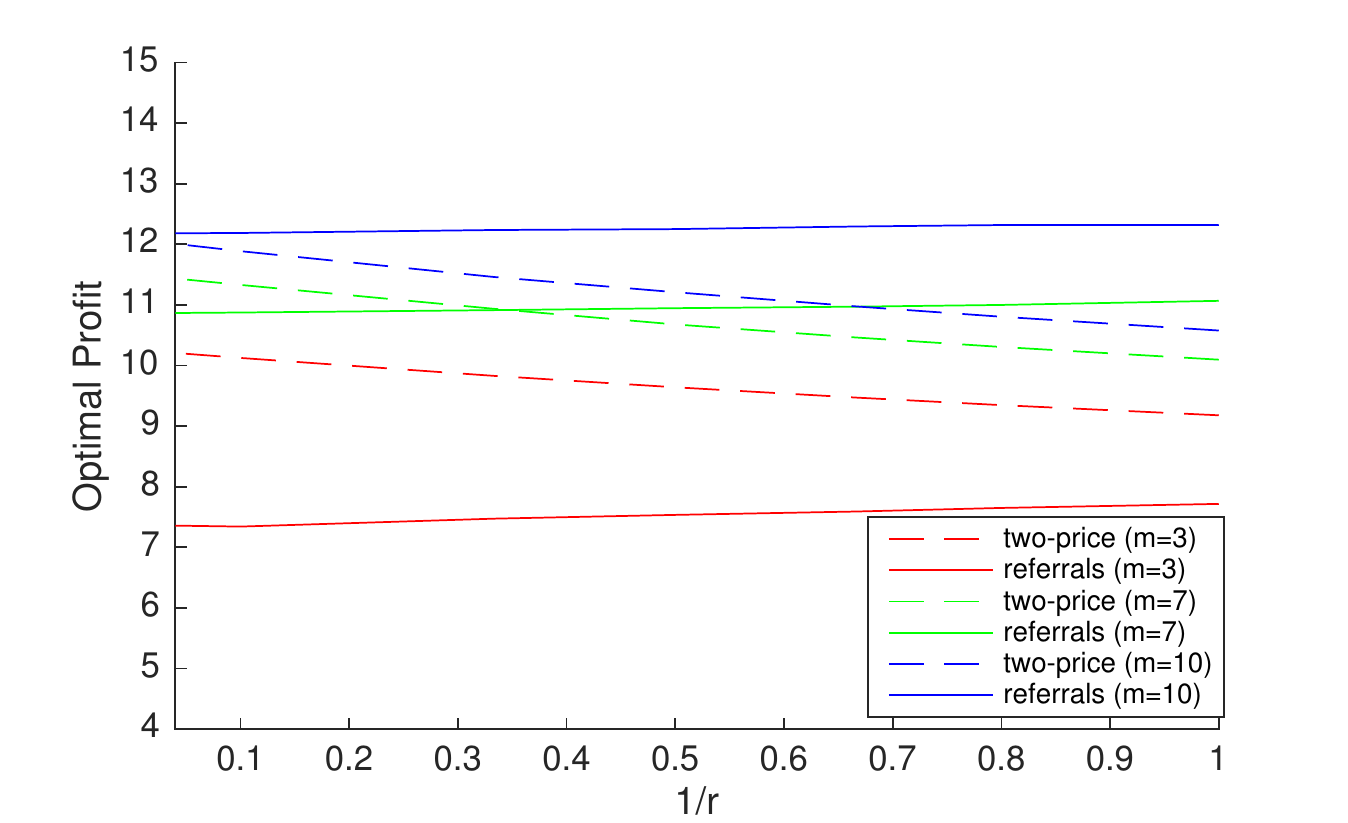}
}
\caption{Optimal Profit of Two-Price (dotted curves) and Referral Incentive (full curves) Policies vs $1/r$. \textit{The degree distributions are as in (\ref{eq:JacksonRogersDistribution}) with the corresponding $r$ and for different fixed values of $m$ (mean degree). Increasing $1/r$ thus increases the degree variance while keeping the average degree fixed. Other model parameters are as in Fig. \ref{ProfitPlotRefDisc}. For interpretation of the references to color in this figure legend, the reader is referred to the online version of this article.}}
\label{r_comp_Stat}
\end{figure*}

To better see the role of the variance of the degree distribution, we vary the parameter $r$. In Figure \ref{r_comp_Stat}, we plot the optimal profit for both classes of pricing policies against $1/r$ varied in the range $[0,1]$. A higher $1/r$ means that we increase the spread of the degree distribution while keeping the mean degree (corresponding to the parameter $m$) constant. This implies that we increase the degree variance while keeping the average degree fixed.
We see that referral policies (full lines) tend to fare better as the variance increases. This creates sufficient thickness in the upper tail of the degree distribution to allow a small fraction of high-degree agents to be incentivized with referrals and generate higher informational efficiency (since the probability of having a high-degree neighbor $\tilde{f}(d)$ increases for high $d$'s, while $f(d)$ also increases but remains relatively lower).  On the other hand, a two-price policy incentivizes only low-degree agents and the latter become less and less influential as the upper tail thickens, since the probability of having a low-degree neighbor decreases (i.e. $\tilde{f}(d)$ decreases for low $d$). Thus two-price policies (dotted lines) tend to fare worse as the degree variance increases. In the case of a moderate mean degree ($m=7$) this thickening of the upper tail of the degree distribution allows referral policies to start dominating two-price policies at some critical value (here, $1/r \approx 0.35$ or likewise $r \approx 3$).

\

\noindent \textbf{Effect of the Mean of the Degree Distribution}

\begin{figure*}
\centerline{
\includegraphics[scale=0.7]{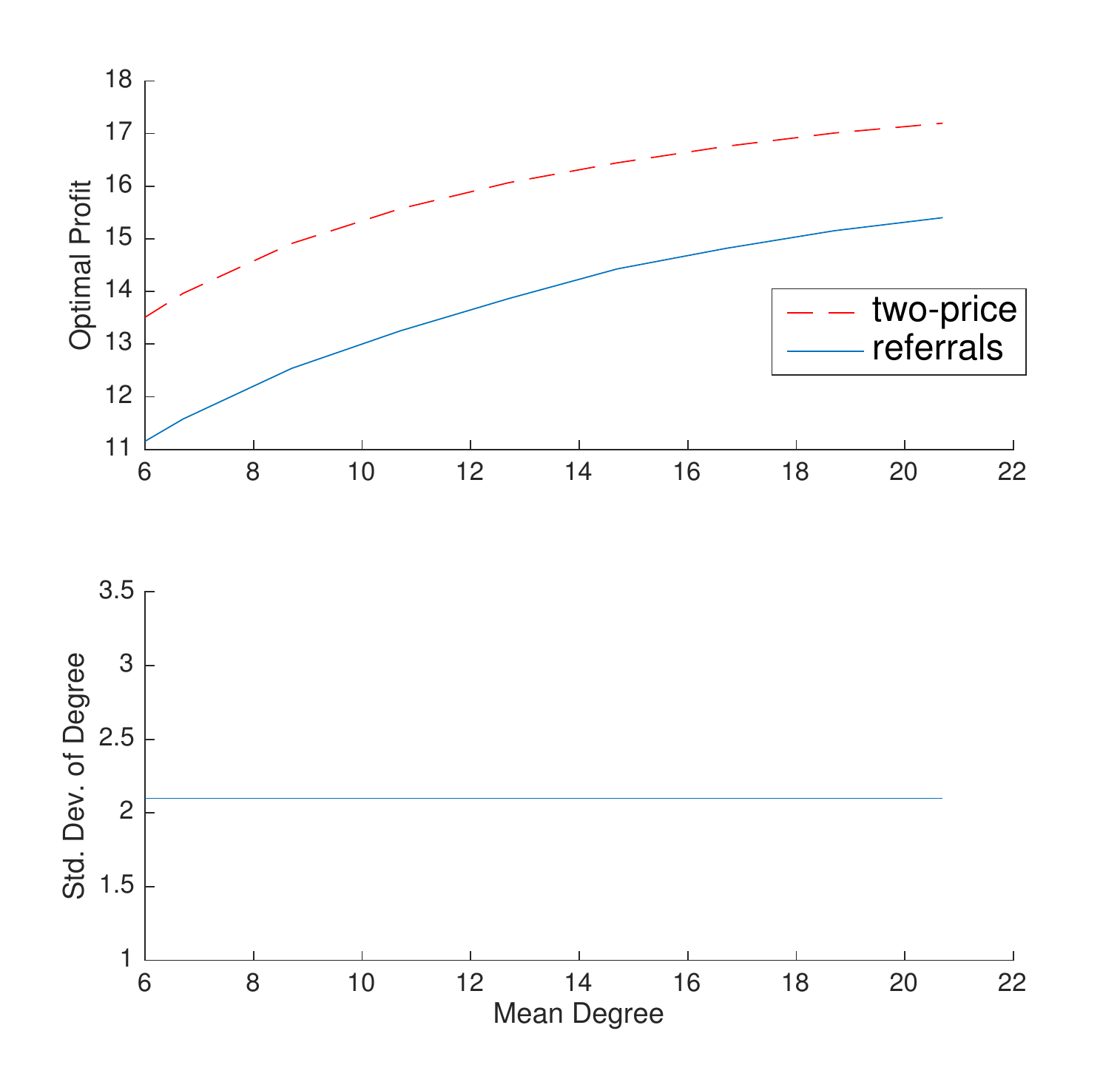}
}
\caption{Optimal Profit of Two-Price and Referral Incentive Policies on a Shifted Two-Degree Distribution. \textit{The degree distributions  are $f(d_u)=0.1$ and $f(d_l)=0.9$, where $d_u = d_l+7$ and $d_l$ is varied from $5$ to $20$. This increases the mean degree while keeping the degree variance fixed. The top panel shows the optimal profits versus the mean degree while the bottom panel shows the standard deviation of the degree versus the mean degree. Model parameters are as in Fig. \ref{ProfitPlotRefDisc}. }}
\label{ProfitVariance_2deg_shifted}
\end{figure*}

To examine the effect of increasing the average degree while keeping the variance fixed, we cannot use the degree distribution in (\ref{eq:JacksonRogersDistribution}). Instead,  we examine a two-degree distribution with $f(d_u)=0.1$ and $f(d_l)=0.9$, where $d_u = d_l+7$ and where $d_l$ is varied from $5$ to $20$. Optimal profits for both referrals and two-price policies are shown in the upper panel of Fig. \ref{ProfitVariance_2deg_shifted}, while the fixed standard deviation of the degree distribution is shown in the lower panel.

Shifting the distribution upwards benefits both types of policies, since the population has fewer low-degree agents likely to remain uninformed under either type of policy. On the other hand, merely increasing the mean degree does not in general allow referrals to achieve higher profits than two-price policies. What really allows a referral policy to dominate a two-price policy is when there is a small fraction of very high-degree agents (relative to the others). In such a case, anyone is likely to be connected to them (because of the high $\tilde{f}(d)$ of such highly influential consumers) and thus the monopolist  only needs to incentivize them with referrals in order to inform the rest of the population. Here, the constant degree spread prevents such a small fraction of highly influential agents from arising. That would only happen if the neighbor probability $\tilde{f}(d)$ became sufficiently high with respect to $f(d)$ for high $d$'s.

\begin{figure*}
\centerline{
\includegraphics[scale=0.65]{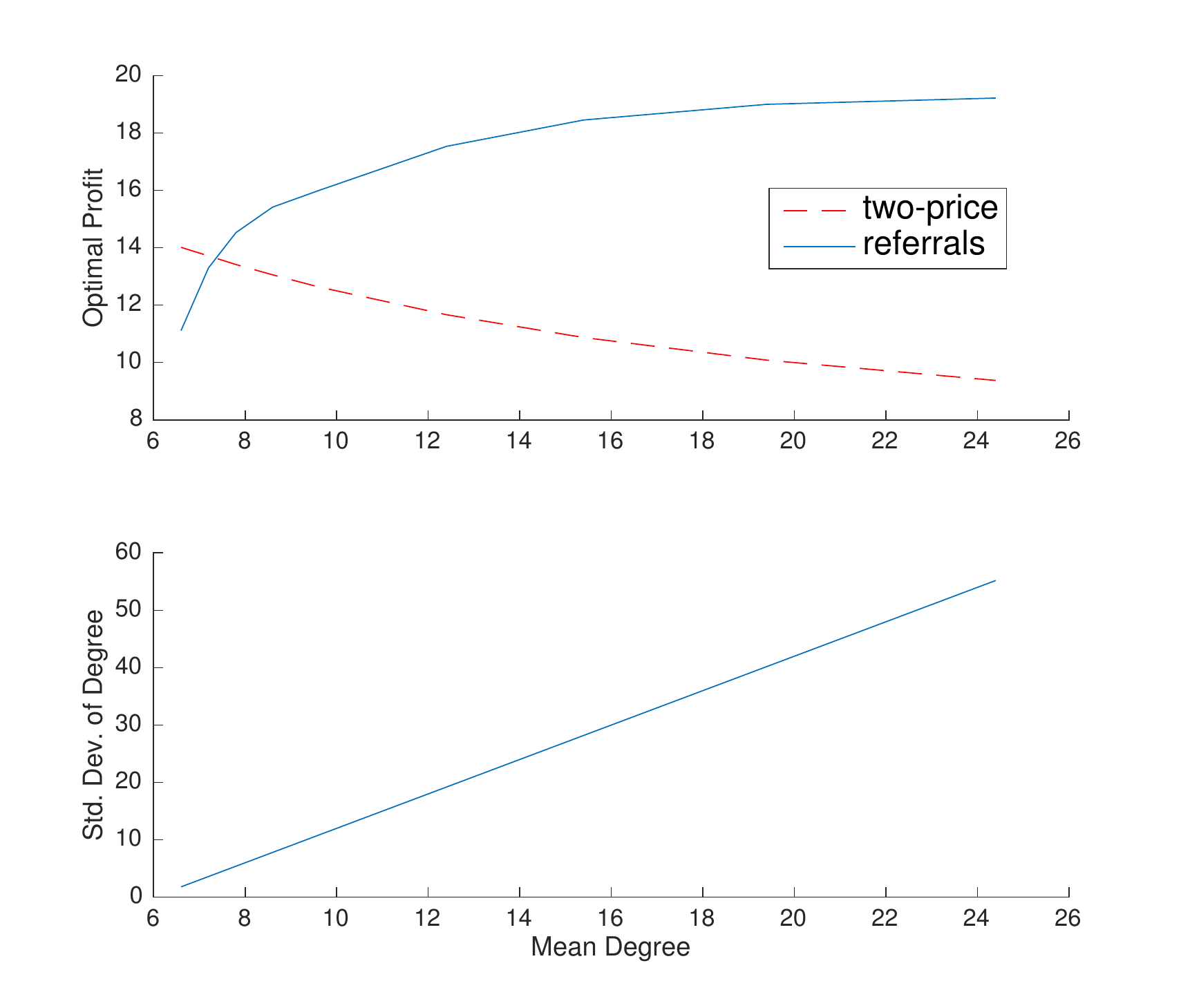}
}
\caption{Optimal Profit of Two-Price and Referral Incentive Policies on a Two-Degree Distribution with Increasing Mean and Standard Deviation. \textit{The degree distributions  are $f(d_u)=0.1$ and $f(d_l)=0.9$, where $d_l = 6$ while $d_u$ is varied from $12$ to $190$. This increases the mean and the standard deviation of the degree distribution. The top panel shows the optimal profits versus the mean degree while the bottom panel shows the standard deviation of the degree versus the mean degree. Model parameters are as in Fig. \ref{ProfitPlotRefDisc}.}}
\label{ProfitVariance_2deg_dHvaried}
\end{figure*}

To illustrate how this works, we examine the effect of increasing the average degree while also increasing the variance. We again use a two-degree distribution with $f(d_u)=0.1$ and $f(d_l)=0.9$, but we now keep $d_l$ fixed at $6$ while we vary $d_u$  from $12$ to $190$. Optimal profits for both referrals and two-price policies are shown in the upper panel of Fig. \ref{ProfitVariance_2deg_dHvaried}, while the standard deviation of the degree is shown in the lower panel. Now we see that the thickening upper tail of the degree distribution can allow referrals to dominate two-price policies, since referrals can now incentivize a small fraction of very influential agents (i.e. agents with high $\tilde{f}(d_u)$) to adopt early and thus to spread information efficiently. Note also that two-price policies are hurt by this increase in the degree spread since low-degree agents are less and less influential (i.e. $\tilde{f}(d_l)$ decreases) and thus such a policy becomes less informative.

\subsection{Welfare}

So far, we have analyzed the effect of different incentive policies on a monopolist's profits.
A regulator may care about social welfare.  Also, in other contexts, a government may want to maximize the welfare associated with the adoption of a program deemed to be of some social value (e.g. health, education, sanitation, more sustainable forms of energy, etc.). It is thus also important to interpret the implications of a pricing policy in terms of welfare.

From a social welfare perspective, given the prior of the agents, any fully optimal allocation would be to have as few agents as
possible experiment and then inform the other agents who would consume only in the second period,\footnote{Given that transfers to the monopolist do not affect welfare, if we account for those profits, then it is only the consumption value of the agents that needs to be considered.}  presuming that information can only flow via the network.\footnote{The social optimum becomes trivial otherwise, as just one agent could experiment and then publicize the results to all other agents.}
This generally favors referral incentives rather than discount pricing, as they encourage experimentation by higher-degree agents rather than low-degree agents. The main caveat is that we have presumed that the monopolist is sure that the product is of high quality. This may lead her to encourage experimentation even when it would not be socially optimal from the prior probability perspective. That is, in some cases in which the prior is so low, it would not make sense to have any agents experiment.   Outside of those cases, the social optimum would be to have a select number of high-degree agents experiment in the first period.
In the case of Proposition \ref{pr:2d_limits} part (i), referral incentives are clearly approximately optimal.
More generally, when there are many high-degree agents (as in Proposition \ref{pr:2d_limits}, parts (iii) and (iv)), then the social optimum would be to ration the referral incentives so that not all of the high-degree agents experiment, but instead, just enough do to inform the other agents.

\subsection{Model Extensions}

\subsubsection{Un-informed Monopolist}

So far we have examined the case where the monopolist is informed of the product quality. If the monopolist is un-informed and has prior $p$ about the quality being high, then two-price policies would no longer dominate referrals on a $d$-regular network. Indeed both referrals ($\mc{R}$-policies) and two-price policies ($\mc{D}$-policies) would yield equal expected profit, which is equal to the optimal expected profit over the whole space of pricing policy triplets $(P_0,P_1,\eta)$. Indeed, referrals are no longer a more expensive form of incentives than two-price policies since it is no longer certain that they will have to be paid out. This is summarized in the following corollary to Theorem \ref{th:discount_opt_d_reg}.

\begin{corollary}[Optimal profit on d-regular networks]
\label{cor:opt_prof_d_reg_uninformed_monop}
Suppose the network is $d$-regular, i.e., $f(d) = 1$ for some $d$ and $f(d)=0$ otherwise. When the monopolist is un-informed about the product quality $\theta$ and shares the prior $p$ with the consumers, then:
\begin{itemize}
\item[(i)] For all $d$, $\hat{\pi}_{\mathcal{D}} = \hat{\pi}_{\mathcal{R}} = \hat{\pi} $;
\item[(ii)] $\lim_{d \to \infty} \hat{\pi}_{\mathcal{D}} = \lim_{d \to \infty}  \hat{\pi}_{\mathcal{R}}=\lim_{d \to \infty} \hat{\pi} = pA_1^H$.
\end{itemize}
\end{corollary}
Note also that in Part (ii), the limiting optimal expected profits now equal $pA^H_1$ instead of $A^H_1$, reflecting the monopolist's prior $p$ about the quality being high.

On the other hand, the results of Proposition \ref{pr:2d_limits} for $2$-degree networks still hold when the monopolist is un-informed. Indeed, because they can incentivize only the high-degree agents, referrals still allow a vanishing fraction of very high-degree consumers to adopt early and inform all other consumers. Note that this is still a profit-maximizing adoption pattern since from Assumption \ref{as:AlwaysFreeriding}, $pA^H_1 > \bar{A}$ and thus the un-informed monopolist prefers late adoption to early adoption. The only difference is that in Part (iv), the optimal expected profit $\hat{\pi}_{\mc{R}}$ is bounded by $pA^H_1$ instead of $A^H_1$, again reflecting the monopolist's prior $p$ about the quality being high.



\subsubsection{Nonlinear Referral Payments}
 Throughout, we have focused on linear referral payments, in which the expected referral payoff scales linearly with an agent's degree. \cite{lobel} study nonlinear referral payments, namely policies that pay up to a maximum number of referrals. Here the monopolist could choose not only the value of the referral payment $\eta$, but also the maximum number of referrals $d^{max} \in \mathbbm{N} \bigcup \{\infty\}$ that a consumer can obtain.  This additional degree of freedom immediately implies that the monopolist can achieve profits that are at least as high as in the case of linear referral payments  -- the latter being only a particular case with $d^{max} = \infty$.

More generally, in our model, nonlinear payments could create a situation where there would be three early adoption thresholds instead of two -- as in the case of linear referrals. Indeed, agents with degrees much higher than $d^{max}$ may have lower incentives to adopt early than agents with degrees just above $d^{max}$, since the former have a higher chance to collect information from neighbors than the latter and thus may prefer delaying adoption. This could create a situation where `low-degree' agents and a separate group of `middle-degree' agents would adopt early, while the others would delay adoption.

However, it is easy to derive conditions under which the two-threshold adoption pattern (Theorem \ref{th:DoubleThreshEq} and the associated properties in Proposition \ref{cor:LowerThreshEq}) would hold with nonlinear referral payments. Indeed, since the probability of collecting information from a neighbor is bounded above by $1$, the benefits of delaying adoption are also bounded above. Thus, if $\eta$ is chosen to be large enough, then early adoption becomes worthwhile for all agents with degrees above a certain threshold. The situation described previously (three adoption thresholds) can then no longer emerge.

\begin{figure*}
\centerline{
\includegraphics[scale=0.8]{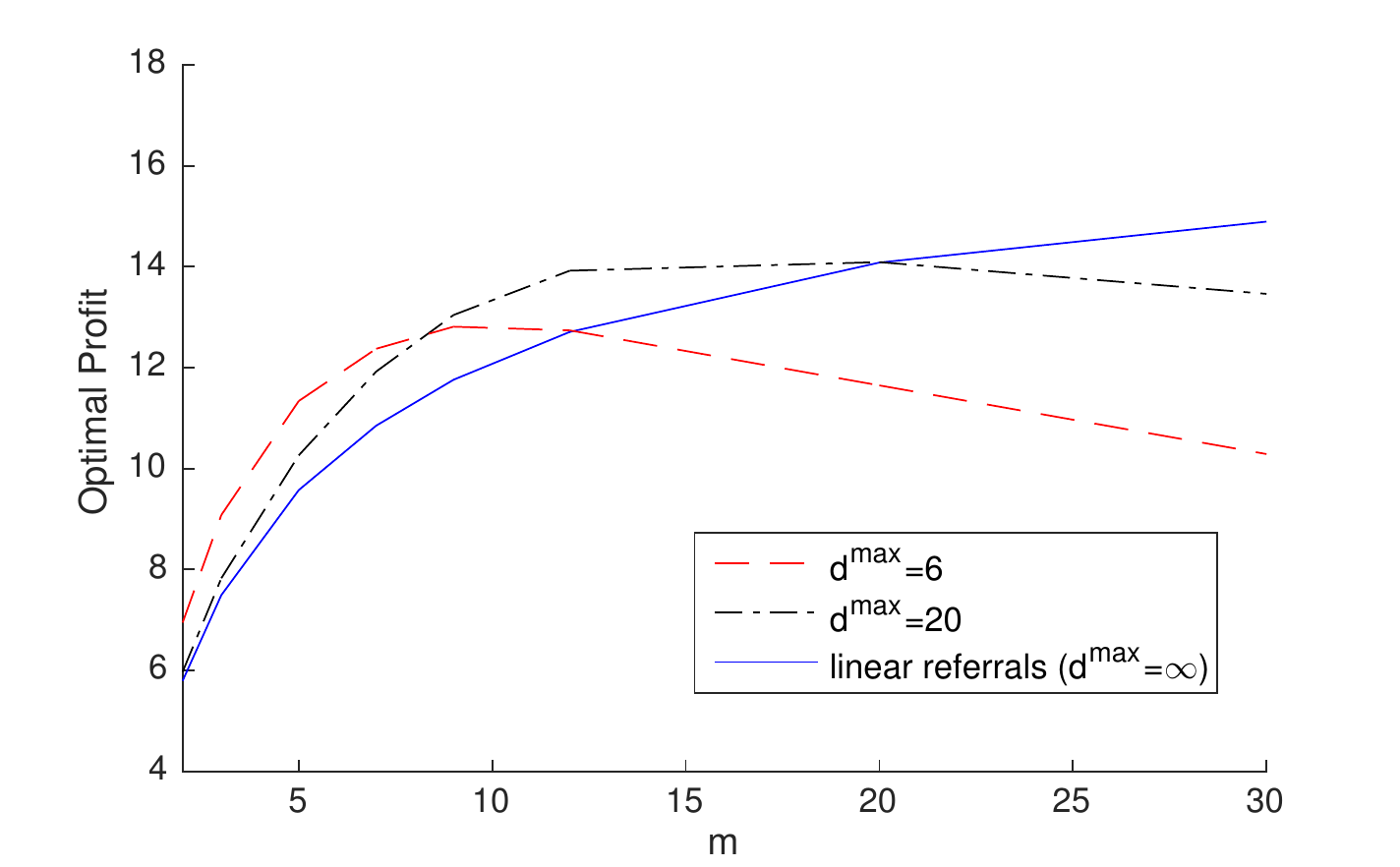}
}
\caption{Optimal Profit of Nonlinear Referral Policies when only a Maximum Number of Referrals Can Be Given. \textit{The degree distributions are as in (\ref{eq:JacksonRogersDistribution}) with $r=2$. The average degree $m$, is varied. Model parameters are as in Fig. \ref{ProfitPlotRefDisc}.}}
\label{non_linear_refs}
\end{figure*}

It is interesting to see how the profit would fare if $d^{max}$ were fixed and the monopolist then optimized over the other parameters $P$ and $\eta$. For illustration purposes, in Fig. \ref{non_linear_refs} we compare the profit using such nonlinear referrals payments to the linear case for different degree distributions (as $m$ is varied). Throughout, $d^{max}$ is held fixed at different values. We see that a lower $d^{max}$ may allow the monopolist to achieve higher profits by reducing the number of referrals that must be paid out. This is particularly true on degree distributions with a low $m$, where the upper-threshold strategy has a relatively low $d_U^*$. In such cases, linear referrals would pay a large mass of higher-degree agents far more referrals than they would require in order to adopt early. On the other hand, a low $d^{max}$ may prevent the monopolist from incentivizing only very high-degree agents and thus achieving a very high $d_U^*$, which is the most efficient adoption strategy on degree distributions with a high $m$. Therefore, while a fixed $d^{max}$ may reduce the cost of incentivizing higher-degree agents, it may also limit the monopolist's ability to price discriminate.

\section{Conclusion}
\label{Conclusion}

When a new product or technology is introduced, there is uncertainty about its quality. This quality can be learned by trying the product, at a risk. It can also be learned by letting others try it and then free-riding on the information that they generate.  The class of network games we developed enabled us to study how choices of agents depend on their degree, finding that agents at both ends of the degree support may choose to adopt early while the rest elect to free-ride on the information generated by the former.

The insights obtained from our analysis also instruct a marketer or monopolist on the design of an optimal dynamic pricing policy. We focused our attention on simple and prominent pricing policies: (i) inter-temporal price discrimination (i.e., early-adoption price discounts) and (ii) referral incentives. We show that the former constitutes a screening mechanism under which lower-degree agents choose to adopt early, whereas the latter constitute a screening mechanism under which higher-degree agents choose to adopt early. The only network information needed to implement such mechanisms is the degree distribution.  We also showed that if agents have the same propensity to interact with each other (modeled by a $d$-regular network), inter-temporal price discrimination yields optimal profits; while if a fraction of agents have a disproportionately large propensity to interact with others, referral incentives yield optimal profits. With richer heterogeneity in degrees, both inter-temporal price discrimination \textit{and} referral incentive policies can be part of profit maximizing choices.

We also showed that in the case where agents have the same propensity to interact with each other, both referrals and inter-temporal price discrimination can be welfare maximizing; whereas when some agents have a disproportionately large propensity to interact with others, referrals yield strictly greater welfare than inter-temporal price discrimination.

This work can help develop more sophisticated models of technology adoption taking into account uncertainty about product quality as well as informational free-riding and the role of referral incentives, as well as more complicated industry structures.\footnote{See \cite{chen} for a discussion of a duopoly in a network product setting.   Their model differs on most dimensions from that here, so
would not apply directly, but examines how incentives of firms differ when they are monopolists versus duopolists.}
One possible future direction would be to endow the monopolist with information about the degrees of consumers and allow her to price discriminate based on that additional information.
Such models can in turn lead to the development of better dynamic pricing mechanisms that optimize the spread of information in large social systems.  This can also inform policy, both in regulating monopolists and also promoting other diffusion processes (e.g., of government or other non-profit programs).

\section{Appendix}

\subsection{Validity of the Mean-Field Approximation}

In this section, we compare the equilibrium outcome in the mean-field setting analyzed so far to the equilibrium outcomes in a finite model, where agents have \underline{full} knowledge of the network topology. Since the second stage ($t=1$) of the game is perfectly determined by the outcome of the first stage ($t=0$), we will only examine the behavior of agents in the first stage. We will compare outcomes under Nash equilibrium and mean-field equilibrium on two types of networks: (i) a complete network and (ii) a star network. We will see that the mean-field model has the advantage of selecting a unique equilibrium outcome and of eliminating implausible equilibria that could arise under full information.

\subsubsection{On A Complete Network:}
Here we assume a set $\mathcal{N}$ of $n$ agents who are fully connected.
\\

\underline{Under a two-price policy (i.e., $\eta=0$):}

There can be many Nash equilibria. There is a symmetric Nash equilibrium in which all agents play a mixed strategy $\mu_i = \omega$. There are also many pure strategy Nash equilibria: $\mu_i=1$ for some $i \in \mathcal{N}$ and $\mu_{-i}=0$ is a Nash equilibrium. There should thus be $n$ such pure-strategy Nash equilibria, in which a single agent experiments and all the others free ride.

An analogous case in the mean-field model is an $n$-regular graph. There is a single symmetric mixed-strategy mean-field equilibrium where all agents play $\mu(n) = \omega$, but this is easily purified to match any of the asymmetric pure strategy equilibria.


\underline{Under a referral policy (i.e., $\eta>\underline{\eta}$):}

There can be many Nash equilibria. There is again a symmetric Nash equilibrium in which all agents play a mixed strategy $\mu_i = \omega$. There may also be many pure strategy Nash equilibria: $\mu_i=1$ for all $i$ in a subset $\mathcal{N}_{\omega}$ of agents such that $\omega = \frac{|\mathcal{N}_{\omega}|-1}{|\mathcal{N}|-1}$ and $\mu_{-i}=0$ is a Nash equilibrium. This is because in such cases all agents are indifferent, i.e.

\begin{equation*}
\Pi^{adopt} = \bar{A} -P + p \eta (1-\omega) (n-1) = p(A_1^H - P) = \Pi^{defer}
\end{equation*}

In the analogous mean-field model on an $n$-regular graph, there is a single symmetric mixed-strategy mean-field equilibrium where all agents play $\mu(n) = \omega$. Thus again, the mean-field model selects the symmetric mixed-strategy equilibrium, but again is easily purified to match any asymmetric pure strategy equilibrium.

\subsubsection{On A Star Network:}
Here we assume a set $\mc{N}$ of $n$ agents: one agent at the center of the star and $n-1$ agents in the periphery. We label the center agent with $i=n$.

\underline{Under a two-price policy (i.e., $\eta=0$):}

There can be many pure and mixed-strategy Nash equilibria. The pure strategy Nash equilibria are $\mu_n = 1$ and $\mu_{-n}=0$ as well as $\mu_n = 0$ and $\mu_{-n}=1$. In other words, either the center experiments and the periphery free-ride or the other way around.

There can be a mixed-strategy Nash equilibrium in which $\mu_n = \omega$ and $\mu_{-n}=0$ and one in which $\mu_n = 0$ and $\mu_{-n}=\omega'$. There can also be a mixed-strategy Nash equilibrium in which $\mu_n = \omega$ and $\mu_{-n}=\omega'$, in other words, the mixed strategy of the center node makes the periphery indifferent and at the same time, the mixed-strategy of the periphery makes the center node indifferent.

The two-degree mean-field model (with $d_l=1$ and $d_u=n-1$) may be understood as somewhat analogous to the star network, although a star is only one possible realization of this two-degree model. In the two-degree case, and depending on the game parameters, the unique mean-field equilibrium can take the form $\mu(n-1)=0$ and $\mu(1)=\omega$, $\mu(n-1)=0$ and $\mu(1)=1$ or  $\mu(n-1)=\omega$ and $\mu(1)=1$, depending on game parameters. These are all lower-threshold strategies. So the mean-field equilibrium effectively selects the equilibrium in which the periphery adopts early and the center free-rides.

\underline{Under a referral policy (i.e., $\eta>\underline{\eta}$):}

We outline here two possible pure-strategy Nash equilibria: with high enough $\eta$, under the finite model, both $\mu_n = 1$ and $\mu_{-n}=0$ as well as $\mu_n = 0$ and $\mu_{-n}=1$ are pure-strategy Nash equilibria. There may also be mixed-strategy Nash equilibria.

Under the mean-field model, with high enough $\eta$, we have an upper-threshold strategy: the unique mean-field equilibrium can have the form $\mu(n-1)=\omega$ and $\mu(1)=0$, $\mu(n-1)=1$ and $\mu(1)=0$ or  $\mu(n-1)=1$ and $\mu(1)=\omega$, depending on game parameters. Thus the mean-field equilibrium effectively selects the equilibrium in which the center adopts early and the periphery free-rides.

\subsection{Effect of Degree Dependence Between Neighbors on Equilibrium Outcomes}

In the mean-field model analyzed so far, the distribution of a  neighbor's degree was given by the edge-perspective degree distribution $\tilde{f}(d)$. It would be interesting to examine how degree dependence between neighbors would affect equilibrium outcomes. For that purpose, let us introduce the following terminology.

Let $\mathbbm{P}(d|k)$ be the probability that a neighbor has degree $d$ when the agent has degree $k$. Then we have the following definition.

\begin{definition}{(Neighbor affiliation).}
We say that the interaction structure exhibits negative (positive) neighbor affiliation if $\mathbbm{P}(d|k) \succeq (\preceq) \ \mathbbm{P}(d|k')$ for $k'>k$, where $\succeq$ indicates first-order stochastic dominance.
\end{definition}

Effectively, \em negative neighbor affiliation \em means that a higher-degree agent is more likely to be connected to a lower-degree agent while \em positive neighbor affiliation \em means that he is more likely to be connected to another higher-degree agent.

Under a two-price policy resulting in a lower-threshold strategy, negative neighbor affiliation would preserve the lower-threshold nature of the equilibrium. Indeed, a lower-degree agent is now more likely to be connected to a high-degree free-rider and thus has an even greater incentive to adopt early. Likewise, a higher-degree free-rider has a higher chance of being connected to a lower-degree early adopter and thus has an even higher incentive to free-ride.

Similarly, under an upper-threshold referral policy, negative neighbor affiliation would preserve the upper-threshold nature of the equilibrium. Indeed, a lower-degree free-rider is now more likely to be connected to a high-degree early adopter and thus has an even greater incentive to free ride. Likewise, a higher-degree early adopter has a higher chance of being connected to a lower-degree free-rider and thus has an even higher incentive to adopt early.

For opposite reasons, positive neighbor affiliation would tend to weaken the threshold results derived in the mean-field model. For example, in a lower-threshold strategy, a lower-degree early adopter is more likely to be connected to another lower-degree early adopter and thus has a smaller incentive to adopt early.

\subsection{Proofs}
\begin{proof}[Theorem \ref{th:existence}]
For any $\alpha \in [0,1]$ define the correspondence $\Phi$ by $\Phi(\alpha) = \alpha(\mc{BR}(\alpha))$.  Any fixed point $\alpha^*$ of $\Phi$, with the corresponding $\mu^* \in \mc{BR}(\alpha^*)$ such that $\alpha(\mu^*) = \alpha^*$ constitute a mean-field equilibrium. We thus need to show that the correspondence $\Phi$ has a fixed point. We employ Kakutani's fixed point theorem on the composite map $\Phi(\alpha) = \alpha(\mc{BR}(\alpha))$.

Kakutani's fixed point theorem requires that $\Phi$ have a compact domain, which is trivial since $[0,1]$ is compact.  Further, $\Phi(\alpha)$ must be nonempty; again, this is straightforward, since both $\mc{BR}(\cdot)$ and $\alpha(\cdot)$ have nonempty image.

Next, we show that $\Phi(\alpha)$ has a closed graph.  We first show that $\mc{BR}$ has a closed graph, when we endow the set of mean-field strategies with the product topology on $[0,1]^\infty$.  This follows easily: if $\alpha_n \to \alpha$, and $\mu_n \to \mu$, where $\mu_n \in \mc{BR}(\alpha_n)$ for all $n$, then $\mu_n(d) \to \mu(d)$ for all $d$.  Since $\pidefer(\alpha, d)$ and $\piadopt(\alpha, d)$ are continuous, it follows that $\mu(d) \in \mc{BR}(\alpha)$, so $\mc{BR}$ has a closed graph.  Note also that with the product topology on the space of mean-field strategies, $\alpha(\cdot)$ is continuous: if $\mu_n \to \mu$, then $\alpha(\mu_n) \to \alpha(\mu)$ by the bounded convergence theorem.

To complete the proof that $\Phi$ has a closed graph, suppose that $\alpha_n \to \alpha$, and that $\alpha'_n \to \alpha'$, where $\alpha'_n \in \Phi(\alpha_n)$ for all $n$.  Choose $\mu_n \in \mc{BR}(\alpha_n)$ such that $\alpha(\mu_n) = \alpha_n'$ for all $n$.  By Tychonoff's theorem, $[0,1]^\infty$ is compact in the product topology; so taking subsequences if necessary, we can assume that $\mu_n$ converges to a limit $\mu$.  Since $\mc{BR}$ has a closed graph, we know $\mu \in \mc{BR}(\alpha)$.  Finally, since $\alpha(\cdot)$ is continuous, we know that $\alpha(\mu) = \alpha'$.  Thus $\alpha' \in \Phi(\alpha)$, as required.

Finally, we show that the image of $\Phi$ is convex.  Let $\alpha_1, \alpha_2 \in \Phi(\alpha)$, and let $\hat{\alpha} = \delta \alpha_1 + (1 - \delta) \alpha_2$, where $\delta \in (0,1)$.  Choose $\mu_1 \in \mc{BR}(\alpha_1)$ and $\mu_2 \in \mc{BR}(\alpha_2)$, and let $\hat{\mu} = \delta \mu_1 + (1- \delta) \mu_2$; note that $\hat{\mu} \in \mc{BR}(\alpha)$ since $\mc{BR}(\alpha)$ is convex.  Finally, since $\alpha(\cdot)$ is linear, we have $\alpha(\hat{\mu}) = \hat{\alpha}$, which shows that $\hat{\alpha} \in \Phi(\alpha)$---as required.

By Kakutani's fixed point theorem, $\Phi$ possesses a fixed point $\alpha^*$.  Letting $\mu^* \in \mc{BR}(\alpha^*)$ be such that $\alpha(\mu^*) = \alpha^*$, we conclude that $\mu^*$ is a mean-field equilibrium.

Next, we prove the second conclusion of the theorem about the uniqueness of $\alpha(\mu^*)$.

We proceed in a sequence of steps, recalling (\ref{DeltaPi}):

{\em Step 1: For all $d \geq 1$, $\piadopt(\alpha,d) - \pidefer(\alpha, d)$ is strictly decreasing in $\alpha \in (0,1)$.}  Note that $\eta \geq 0$ and $p > 0$, so it follows from \eqref{eq:mf_adopt_t0} that $\piadopt(\alpha,d)$ is non-increasing in $\alpha$.  Further, $A_1^H > 0$ so it follows from \eqref{eq:mf_delay_t0} that $\pidefer(\alpha, d)$ is strictly increasing in $\alpha$, as required.

{\em Step 2: For all $d \geq 1$, and $\alpha' > \alpha$, $\mc{BR}_d(\alpha') \preceq \mc{BR}_d(\alpha)$.}\footnote{Here the set relation $A\preceq B$ means that for all $x \in A$ and $y \in B$, $x \leq y$.}  This follows immediately from Step 1 and the definition of $\mc{BR}_d$ in Section \ref{sec:BestResponse}.

{\em Step 3: If $\mu'$, $\mu$ are mean-field strategies such that $\mu(d) \geq \mu'(d)$, then $\alpha(\mu) \geq \alpha(\mu')$.}  This follows since $\alpha(\cdot)$ is linear in its arguments, with nonnegative coefficients.

{\em Step 4: Completing the proof.}  So now suppose that there are two mean field equilibria $(\mu^*,\alpha^*)$ and $(\mu'^{*},\alpha'^{*})$, with $\alpha'^{*} > \alpha^*$.  By Step 2, since $\mu^* \in \mc{BR}(\alpha^*)$ and $\mu'^{*} \in \mc{BR}(\alpha'^{*})$, we have $\mu^*(d) \geq \mu'^{*}(d)$.  By Step 3, we have $\alpha^* =\alpha(\mu^*) \geq \alpha(\mu'^{*}) = \alpha'^{*}$, a contradiction.  Thus the $\alpha^*$ in any mean-field equilibrium must be unique. It follows that $\mu^*$ is unique in this sense, as required.
\end{proof}

\begin{proof}[Theorem \ref{th:DoubleThreshEq}]
Consider now, $\Delta  \Pi(\alpha,d)$ as a function of the continuous variable $d$ over the connected support $[1,\infty)$. First note that (\ref{DeltaPi}) can be rewritten as
\begin{eqnarray}
\Delta \Pi(\alpha,d) &=& \Pi^{adopt}(\alpha,d) - \Pi^{defer}(\alpha,d)  \nonumber \\
&=&  p A^H_0 + (1-p)A_0^L + \eta p (1- \alpha)d + p A_1^H  (1 - \alpha)^d \nonumber
\end{eqnarray}
For any $\alpha \in (0,1)$, $\Delta  \Pi(\alpha,d)$ is the sum of a non-decreasing affine function of $d$
and a convex function of $d$. $\Delta  \Pi(\alpha,d)$ is therefore convex in $d$. It follows that it is
also quasiconvex and thus the inverse image of $(-\infty,0)$ is a convex set, specifically, an interval
$[1,y)$ if $\Delta  \Pi(\alpha,1) < 0$ or an interval $(x,y)$ where $x \geq 1$, otherwise. The integers in
such intervals (i.e., $[1,y) \bigcap \mathbbm{N}^+$ or $(x,y) \bigcap \mathbbm{N}^+$) represent the
degrees of the agents for whom delaying adoption is a strict best response, i.e.,
$\{d: \mc{BR}_{d}(\alpha) = \{0\}\}$. It follows that the degrees of agents for whom early
adoption is a strict best response, i.e., $\{d: \mc{BR}_{d}(\alpha) = \{1\} \}$, are
located outside of this interval, i.e., at either or both extremities of the degree support.
This result holds for any couple $(\mu,\alpha)$, where $\mu \in \mc{BR}(\alpha)$ and $\alpha \in (0,1)$. It therefore holds for any mean-field equilibrium $(\mu^*,\alpha^*)$ such that $\alpha^* \in (0,1)$.

Note that any mean-field equilibrium $\mu^*$ has the same corresponding  $\alpha^* = \alpha(\mu^*)$ (cf. Theorem \ref{th:existence}). Letting
$d_L^* = \sup \{ z: \mc{BR}_d (\alpha^*) = \{1\}, \text{ for all } d<z \}$\footnote{Note that in the event that $\mc{BR}_{d}(\alpha^*) \neq \{1\} $ for $d=1$, we set $d^*_L = 1$.}
and
$d_U^* = \inf \{ z: \mc{BR}_d (\alpha^*) = \{1\}, \text{ for all } d>z \}$ 
defines a pair of thresholds $d_L^*$ and $d_U^*$ valid for all strategies that may arise in a mean-field
equilibrium, i.e., any $\mu^*$ such that $\mu^* \in \mc{BR}(\alpha^*)$ and $\alpha^* = \alpha(\mu^*)$. 
\end{proof}

\begin{proof}[Proposition \ref{cor:LowerThreshEq}]
{\em Part (1):} We first prove the existence of $\underline{\eta}$ in the first part of the proposition. We give a sufficient condition. Let us show that there exists some $\underline{\eta}'$ such that $\forall \eta < \underline{\eta}'$, $\Delta \Pi(\alpha,\bar{d}-1) > \Delta \Pi(\alpha,\bar{d})$ for all $\alpha \in (0,1)$. First note that using (\ref{DeltaPi}), we can write
\begin{equation}
\Delta \Pi (\alpha,\bar{d}) - \Delta \Pi (\alpha,\bar{d}-1) =  \eta p (1-\alpha) - \alpha pA_1^H(1-\alpha)^{\bar{d}-1}
\label{eq:DeltaPi_lower_th}
\end{equation}

For any $\alpha \in (0,1)$, if $\eta <  \frac{\alpha pA_1^H(1-\alpha)^{\bar{d}-1} }{p(1-\alpha)}$, the RHS of (\ref{eq:DeltaPi_lower_th}) is negative and thus $\Delta \Pi (\alpha,\bar{d}) < \Delta \Pi (\alpha,\bar{d}-1)$. Note that this is obviously also true when $\eta=0$ and remains so for small positive values of $\eta$ (i.e., $\eta<\underline{\eta}'$) by continuity (for any $\alpha \in (0,1)$, given any finite $\bar{d}$ and for some $\underline{\eta}'>0$). By the convexity of $\Delta \Pi (\alpha,d)$ in $d$ (cf. proof of Theorem \ref{th:DoubleThreshEq}), it follows by induction that $\Delta \Pi (\alpha,d+1) < \Delta \Pi (\alpha,d)$, for all $1 \leq d < \bar{d}$.

{Note that in any best response $\mu \in \mc{BR}(\alpha)$ to $\alpha$, $\mu(d) = 1$ whenever $\Delta \Pi (\alpha, d) > 0$, and $\mu(d) = 0$ whenever $\Delta \Pi (\alpha, d) < 0$.  Since $\Delta \Pi (\alpha, d)$ is strictly decreasing in $d$, the desired result follows for any $\eta < \underline{\eta}'$  We now set $\underline{\eta}=\sup \{ \underline{\eta}': d^*_U > \bar{d} \}$ so that $\underline{\eta}$ is the largest $\eta$ such that every mean-field equilibrium can be characterized by $d^*_U > \bar{d}$.}

{\em Part (2):} We give a sufficient condition. Let us show that there exists $\bar{\eta}' < \infty$ such that $\forall \eta > \bar{\eta}'$, $\Delta \Pi (\alpha,1) < \Delta \Pi (\alpha,2)$, for all $\alpha \in (0,1)$. First note that using (\ref{DeltaPi}), we can write
\begin{eqnarray}
\Delta \Pi (\alpha,2) - \Delta \Pi (\alpha,1) &=&  \eta p (1-\alpha) - pA_1^H(1-\alpha -(1-\alpha)^2)  \nonumber \\
								&=& p(\eta - A_1^H) - \alpha (p(\eta - A_1^H))
								 + pA_1^H(1-\alpha)^2 \nonumber
\end{eqnarray}
We verify that when {$\eta > \bar{\eta}' = A_1^H$}, $\Delta \Pi (\alpha,2) - \Delta \Pi (\alpha,1) > 0$, for all $\alpha \in (0,1)$. Note that $\Delta \Pi (\alpha,2) - \Delta \Pi (\alpha,1)$ is the sum of an affine function of $\alpha$ and a purely quadratic function of $\alpha$. The quadratic term is strictly positive over the desired range of $\alpha$. The affine term is also strictly positive when $\eta > A_1^H $. It thus follows that $\Delta \Pi (\alpha,2) > \Delta \Pi (\alpha,1)$ for any $\alpha \in (0,1)$, when $ \eta > \bar{\eta}'$.

By the convexity of $\Delta \Pi (\alpha,d)$ in $d$ (cf. the proof of Theorem \ref{th:DoubleThreshEq}), it follows by induction that $\Delta \Pi (\alpha,d+1) > \Delta \Pi (\alpha,d)$, for all $d \geq 1$. {Since $\Delta \Pi (\alpha, d)$ is strictly increasing in $d$, the result follows for $\eta > \bar{\eta}'$.  We now set $\bar{\eta}= \inf \{\bar{\eta}' : d_L^*=0\}$.}

{\em Part (3):}. By construction of $\underline{\eta}$ and $\bar{\eta}$ in the previous two parts of the proof, it follows that any $\underline{\eta}<\eta<\bar{\eta}$ leading either to $d^*_L = 0$ or $d^*_U> \bar{d}$ is a contradiction. We thus conclude that  for all $\underline{\eta}<\eta<\bar{\eta}$, any mean-field equilibrium can be characterized by some $d_L^* \geq 1$ and $d^*_U \leq \bar{d}$.

\end{proof}

\begin{proof}[Proposition \ref{pr:FOSD_f_tilde}]
Let $\mu^*$ and $\mu'^*$ be mean-field equilibria arising under the distributions $\tilde{f}$ and $\tilde{f}'$ respectively and let $\alpha^* = \alpha(\mu^*)$ and $\alpha'^* = \alpha(\mu'^*)$.

We start by proving the first part of the proposition.  Suppose $\alpha'^* > \alpha^*$. Then $\mc{BR}(\alpha'^*) \preceq \mc{BR}(\alpha^*)$ (cf. Proof of Theorem \ref{th:existence}, \textit{Step 2}). {Thus we have $\mu'^*(d) \leq \mu^*(d)$ for all $d$, from which we obtain:
\begin{equation}
\label{eq:FOSD_mu_ineq}
\alpha'^* = \sum_{d \geq 1} \tilde{f}'(d) \mu'^*(d) \leq \sum_{d \geq 1} \tilde{f}'(d) \mu^*(d).
\end{equation}
Since $\eta < \underline{\eta}(\tilde{f})$, $\mu^*$ must be a lower threshold strategy, i.e., there exists $d_L^*$ such that $\mu(d) = 1$ for all $d < d_L^*$ and $\mu(d) = 0$ for all $d > d_L^*$.  In other words, $\mu^*$ is a decreasing function.  Since $\tilde{f}'$ first order stochastically dominates $\tilde{f}$ and $\mu^*$ is decreasing, we obtain:
\[ \sum_{d \geq 1} \tilde{f}'(d) \mu^*(d) \leq \sum_{d \geq 1} \tilde{f}(d) \mu^*(d) = \alpha^*, \]
which, when combined with \eqref{eq:FOSD_mu_ineq} yields $\alpha'^* \leq \alpha^*$, a contradiction.  We conclude that $\alpha'^* \leq \alpha^*$, as required.}

{The proof of the second part of the proposition follows in an analogous manner.  We start by assuming that $\alpha'^* < \alpha^*$, so that $\mc{BR}(\alpha'^*) \succeq \mc{BR}(\alpha^*)$, reversing the inequality in \eqref{eq:FOSD_mu_ineq}:
\begin{equation}
\label{eq:FOSD_mu_ineq2}
\alpha'^* = \sum_{d \geq 1} \tilde{f}'(d) \mu'^*(d) \geq \sum_{d \geq 1} \tilde{f}'(d) \mu^*(d).
\end{equation}
Further, since $\mu^*$ is an upper threshold strategy (cf. Proposition \ref{cor:LowerThreshEq}), we conclude that it is increasing, so that:
\[ \sum_{d \geq 1} \tilde{f}'(d) \mu^*(d) \geq \sum_{d \geq 1} \tilde{f}(d) \mu^*(d) = \alpha^*, \]
which, when combined with \eqref{eq:FOSD_mu_ineq2} yields $\alpha'^* \geq \alpha^*$, a contradiction.  We conclude that $\alpha'^* \geq \alpha^*$, as required.}

\end{proof}

\begin{proof}[Proposition \ref{obs:LowerThresholdMech}]
Setting $\eta = 0$ in (\ref{DeltaPi}) and incorporating prices $P_0$ and $P_1$ results in
\begin{equation}
\Delta \Pi(\alpha,d) = p(A_0^H+A_1^H) + (1-p)A_0^L - P_0 - p(A^H_1-P_1)+ p (A_1^H-P_1)  (1 - \alpha)^d \nonumber
\end{equation}
We restrict our attention to the case where $P_1<A^H_1$, else no agent adopts late. Consider again $\Delta \Pi(\alpha,d) $ as a function of a continuous variable $d$ over the connected support $[1,\infty)$. For any $\alpha \in (0,1)$, $\Delta \Pi(\alpha,d) $ is a strictly decreasing function of $d$.  {The result follows as in Part (1) of Proposition \ref{cor:LowerThreshEq}.}

\end{proof}

\begin{proof}[Proposition \ref{obs:UpperThresholdMech}]
{The proof is essentially identical to the proof of Part (2) of Proposition \ref{cor:LowerThreshEq}.}
First note that using (\ref{DeltaPi}) and incorporating prices $P_0$ and $P_1$, we can write
\begin{eqnarray}
\Delta \Pi (\alpha,2) - \Delta \Pi (\alpha,1) &=&  \eta p (1-\alpha) - p(A_1^H-P_1)(1-\alpha -(1-\alpha)^2)  \nonumber \\
								&=& p(\eta - (A_1^H-P_1)) - \alpha (p(\eta - (A_1^H-P_1)))
								 + p(A_1^H-P_1)(1-\alpha)^2 \nonumber
\end{eqnarray}
We restrict our attention to the case where $P_1<A^H_1$, else no agent adopts late.  {Letting $\eta^+ = A_1^H - P_1$, it follows that if $\eta > A_1^H - P_1$, then $\Delta \Pi(\alpha, 2) > \Delta \Pi(\alpha, 1)$ for any $\alpha \in (0,1)$.}

By the convexity of $\Delta \Pi (\alpha,d)$ in $d$ (cf. proof of Theorem \ref{th:DoubleThreshEq}), it follows by induction that $\Delta \Pi (\alpha,d+1) > \Delta \Pi (\alpha,d)$, for all $d \geq 1$.  {Thus $\Delta \Pi(\alpha, d)$ is strictly increasing in $d$, and the result follows as in Part (2) of Proposition \ref{cor:LowerThreshEq}.}


\end{proof}

\begin{proof}[Proposition \ref{prop:poolingEq}]

First we argue that separating equilibria leading to positive profits for at least one type do not exist.
A separating equilibrium must involve 0 profits for a Low type since consumers will never buy from a known Low type at nonnegative prices.
A strategy that earns positive profits for the High Type has to have a positive price in the first period, otherwise all consumers can get the product at a 0 or lower price in the first period (given that it is
known to be a High type via the separation) and will thus never pay a positive price.   If some consumers buy in the first period at the positive price, then it could not be a separating equilibrium, since the Low type must be getting 0 profits in the separating equilibrium and could instead deviate to mimic the High type
and earn positive profits in the first period, with no sales in the next period (as nobody will buy, given that the second period price is nonnegative and so no referral payments will be made).

Next, let us examine the pooling equilibria.  We show that Bayesian beliefs $\mathbbm{P}\{\theta=H| P_0,P_1,\eta\}$ on the equilibrium path are
consistent with equilibrium play. Let the monopolist play $(\hat{P}_0,\hat{P}_1,\hat{\eta})$ as a
pure strategy when $\theta$ is either $H$ or $L$. Then consumers cannot infer any information
about $\theta$ by observing the pricing policy. It follows that
$\mathbbm{P}\{\theta=H| \hat{P}_0,\hat{P}_1,\hat{\eta} \}=p$, i.e., the posterior equals the prior.

We now show that strategies are sequentially rational. Given $(\hat{P}_0,\hat{P}_1,\hat{\eta})$ being played as a pure strategy by the monopolist, consumers play $\mu^* \in \mc{EQ}(\hat{P}_0,\hat{P}_1,\hat{\eta})$, which is their best response to this pricing policy given their belief $\mathbbm{P}\{\theta=H| \hat{P}_0,\hat{P}_1,\hat{\eta}\}=p$.

For the monopolist, when $\theta=H$, the policy $(\hat{P}_0,\hat{P}_1,\hat{\eta})$ leads to non-negative profits $\pi(\hat{P}_0,\hat{P}_1,\hat{\eta},H)$ by construction. On the other hand, any deviation
$(P_0,P_1,\eta) \neq (\hat{P}_0,\hat{P}_1,\hat{\eta})$ leads to consumers forming the belief
$\mathbbm{P}\{\theta=H| P_0,P_1,\eta\}=0$, which means no adoption (i.e., $\mu^* = 0$) and
thus zero profit, i.e., $\pi(P_0,P_1,\eta,H)=0$. It thus follows that
$\pi(\hat{P}_0,\hat{P}_1,\hat{\eta},H) \geq \pi(P_0,P_1,\eta,H)$ for any $(P_0,P_1,\eta)$.

Likewise, for the monopolist, when $\theta=L$, the policy $(\hat{P}_0,\hat{P}_1,\hat{\eta})$ leads to non-negative profits by construction. On the other hand, any deviation
$(P_0,P_1,\eta) \neq (\hat{P}_0,\hat{P}_1,\hat{\eta})$ leads to consumers forming the belief
$\mathbbm{P}\{\theta=H| P_0,P_1,\eta\}=0$, which means no adoption (i.e., $\mu^* = 0$)
and thus zero profit, i.e., $\pi(P_0,P_1,\eta,L)=0$. It thus follows that
$\pi(\hat{P}_0,\hat{P}_1,\hat{\eta},L) \geq \pi(P_0,P_1,\eta,L)$ for any $(P_0,P_1,\eta)$.

Thus, we have shown sequential rationality.

Then, the monopolist choosing pricing policies $(P_0,P_1,\eta)_H = (P_0,P_1,\eta)_L =(\hat{P}_0,\hat{P}_1,\hat{\eta})$ in each state of the world $\theta \in \{H,L\}$ and consumers playing $\mu^* \in \mc{EQ}(\hat{P}_0,\hat{P}_1,\hat{\eta})$ with posterior belief $\mathbbm{P}\{\theta = H|\hat{P}_0,\hat{P}_1,\hat{\eta}\}=p$ and playing $\mu^*=0$ with $\mathbbm{P}\{\theta = H|P_0,P_1,\eta\}=0$ for any other policy $(P_0,P_1,\eta) \neq(\hat{P}_0,\hat{P}_1,\hat{\eta})$ constitute a Perfect Bayesian Equilibrium$^*$ (PBE$^*$).

\end{proof}

\begin{proof}[Proposition \ref{prop:opt_two_price_pol}]

We will first rule out some prices. Let $P_0 > \bar{A}$. Then $\piadopt(\alpha,d) = \bar{A} - P_0 < 0$ for any $d$ and thus any agent's best response is not to adopt early. This leads to $\alpha = \alpha(\mu) = 0$ and thus there is neither early nor late adoption and thus the profit $\rho(P_0,P_1,0,H)$ necessarily equals 0. We can therefore restrict our attention to $P_0 \leq \bar{A}$. Likewise, $P_1 > A^H_1$ leads to a second-period payoff of $A^H_1 - P_1 < 0$ upon late adoption and thus any agent's best response is not to adopt late. We can therefore restrict our attention to $P_1 \leq A^H_1$.

We will now show that any lower-threshold strategy $\mu$ with some threshold $d_L$ can be sustained as a unique equilibrium by some two-price policy. Let $\alpha = \alpha(\mu)$ be the corresponding neighbor adoption probability. Then for some $P_1 < A^H_1$, we can find a particular $P_0 < \bar{A}$ such that

\begin{equation}
\label{eq:pi_adopt_equal_pi_delay_proof_opt_2_price}
\piadopt(\alpha,d) = \bar{A} - P_0 = p (1 - (1-\alpha)^{d_L}) (A^H_1 - P_1) =\pidefer(\alpha,d)
\end{equation}
With such a couple $P_0,P_1$, a degree-$d_L$ agent is indifferent and agents with degrees $d<d_L$ strictly prefer to adopt early while agents with degrees $d>d_L$ strictly prefer to adopt late. The strategy of agents of degree $d_L$, i.e. $\mu(d_L)$, is then chosen so that $\alpha= \alpha(\mu)$. There is a single such $\mu(d_L)$. We have thus shown that for any lower-threshold strategy $\mu$, there exists a couple $P_0,P_1$ such that $\mu$ is sustained as the unique equilibrium.

Note that from (\ref{eq:pi_adopt_equal_pi_delay_proof_opt_2_price}), $P_0$ is increasing in $P_1$. Indeed, there exist increasing sequences $P_{1,k} \uparrow A^H_1$ and $P_{0,k} \uparrow \bar{A}$ such that $\mu \in \mc{EQ}(P_{0,k},P_{1,k},0)$ is the unique equilibrium strategy for all $k$. To see this let  $P_{1,k} \uparrow  A^H_1$ be any monotonically increasing sequence converging to $A^H_1$. Then, simply set $P_{0,k} = \bar{A} - p (1 - (1-\alpha)^{d_L}) (A^H_1 - P_{1,k})$.  Since $\mc{EQ}(P_{0,k},P_{1,k},0)$ is a singleton set,  from (\ref{eq:fixed_policy_profit}) we may thus write

\begin{equation}
\rho(P_{0,k},P_{1,k},0,H) = P_{0,k} \beta(\mu) + P_{1,k} \mc{\gamma}_H(\mu)
\end{equation}

Note that $\rho(P_{0,k},P_{1,k},0,H)$ is strictly increasing in $k$. Moreover, since we have shown that any lower-threshold strategy $\mu$ can be sustained as the unique equilibrium along an appropriate choice of increasing sequences $P_{1,k} \uparrow A^H_1$ and $P_{0,k} \uparrow \bar{A}$, we may then write

\begin{eqnarray}
\pi_{\mc{D}}(\bar{A},A_1^H,H) &=& \limsup_{P_0' \uparrow \bar{A}, P_1'\uparrow A^H_1 } \rho(P_0', P_1', 0,H) \\
 &=& \max_{\{\mu: d_U=\infty\}} \bar{A} \beta(\mu) + A_1^H \mc{\gamma}_H(\mu) \\
 &>& \max_{\{\mu: d_U=\infty\}} P_0 \beta(\mu) + P_1 \mc{\gamma}_H(\mu) \\
 &=& \limsup_{P_0' \uparrow P_0, P_1'\uparrow P_1 } \rho(P_0', P_1', 0,H) \\
 &=& \pi_{\mc{D}}(P_0,P_1,H)
\end{eqnarray}
for any $P_0<\bar{A}$ and $P_1<A_1^H$. The first equality is just the definition of profit from (\ref{eq:Profit_D}). The second equality follows from the fact that $\lim_{P_{0,k} \uparrow \bar{A}, P_{1,k}\uparrow A_1^H } P_{0,k} \beta(\mu) + P_{1,k} \mc{\gamma}_H(\mu)=\bar{A} \beta(\mu) + A_1^H \mc{\gamma}_H(\mu)$ for  sequences $P_{1,k} \uparrow A^H_1$ and $P_{0,k} \uparrow \bar{A}$ such that $\mu \in \mc{EQ}(P_{0,k},P_{1,k},0)$ for any $k$ and from the fact that any lower threshold strategy $\mu$ can be sustained as the unique equilibrium along an appropriate choice of such sequences.

It then follows that $\hat{\pi}_{\mc{D}}= \max_{P_0,P_1} \pi_{\mc{D}}(P_0,P_1,H)=\pi_{\mc{D}}(\bar{A},A_1^H,H)$.

\end{proof}

\begin{proof}[Theorem \ref{th:discount_opt_d_reg}]
{\em Part (i):}
Note that the profit on a $d$-regular network with an $\alpha$-equilibrium,
using \eqref{eq:fixed_policy_profit}, is:
\begin{equation}
\label{eq:pi_disc_d_reg}
\rho(P_0, P_1, \eta,H) = \alpha (P_0 - \eta(1 - \alpha)d) + (1-\alpha)(1-(1-\alpha)^d) P_1 .
\end{equation}
Next, note that if $\alpha\in (0,1)$, then the indifference of the agents implies that
the expected utility from early adoption equals that from waiting:
\begin{equation}
\label{d_reg_eq_condition}
\bar{A} - P_0 + p \eta (1-\alpha)d = p(1 - (1-\alpha)^d)(A_1^H - P_1).
\end{equation}
Thus, any $P_0,\eta$ combination that solves (\ref{d_reg_eq_condition}) can lead to the same $\alpha$.
Note that from (\ref{eq:pi_disc_d_reg}) and (\ref{d_reg_eq_condition}), whenever $\alpha\in (0,1)$, the
profits are proportional to $P_0 - \eta(1 - \alpha) d$ while the indifference condition
of the agent is linear in $P_0 - p \eta (1-\alpha)d$.  Given that $p\in (0,1)$, it follows that
profit maximization requires that if $\alpha\in (0,1)$, then $\eta=0$.

Now note that for any $(P_0, P_1, \eta)$  where $\alpha = 0$, it follows that profits are 0: $\rho(P_0, P_1, \eta,H) = 0$ (since there are no early adopters, and then nobody adopts
in the second period), and that such profits can be achieved by either sort of policy (either setting $\eta=0$ or $P_0$ very high).
Similarly, if $\alpha=1$, referral incentives are irrelevant since there are no second period adopters and so no referral payments.

These observations imply that any sequence of profits associated with some sequence of $\alpha_k$'s and policies can also always be achieved with 0 referral incentives, and so
$\hat{\pi}_{\mathcal{D}} = \hat{\pi} \geq \hat{\pi}_{\mathcal{R}}$.  From the above observations, we can show that the last inequality is strict
if we show that profit maximization requires an $\alpha\in (0,1)$.

To complete the proof we argue that an optimizing $\alpha$ must lie in $(0,1)$, and from the above we can restrict attention to $\eta=0$.

Note that if $P_1 > A_1^H$, then no agent adopts late, since their payoff is negative in that case.  Thus it follows that for any $P_1 > A_1^H$,
we have $\rho(P_0, P_1, 0,H) \leq \rho(P_0, A_1^H, 0,H)$.
Given the definition of the profit $\pi(P_0, P_1, 0,H)$ in \eqref{eq:FullProfit}, we henceforth assume that $P_1 < A_1^H$.

Next, note that if for a given policy $(P_0, P_1, 0)$, we have $\bar{A} - P_0 \geq p(A_1^H - P_1)$, then by comparing $\Pi^{adopt}(\alpha, d)$ and $\Pi^{defer}(\alpha, d)$ we see that $\alpha = 1$.  In this case the profit of the monopolist is $P_0$.  Since we have already assumed $P_1 < A_1^H$, in this case $P_0 < \bar{A}$.  For the moment we note this fact; below we will show this profit is suboptimal, so that we can ignore the possibility that $\alpha = 1$.

So now assume that $\bar{A} - P_0 < p(A_1^H - P_1)$.  This is the equivalent of Assumption \ref{as:AlwaysFreeriding} for the game with pricing.  In this case, Theorem \ref{th:existence} applies, to ensure that every equilibrium $\mu$ leads to the same $\alpha = \alpha(\mu)$.   On a $d$-regular network, given an equilibrium strategy $\mu \in \mc{EQ}(P_0,P_1,\eta)$,
\begin{equation*}
\beta(\mu) = \mu(d)=\alpha(\mu)=\alpha
\end{equation*}
is the equilibrium fraction of early adopters, and
\begin{equation*}
\gamma_H(\mu) = (1-\mu(d))(1- (1-\alpha)^d)=(1-\alpha)(1- (1-\alpha)^d)
\end{equation*}
is the equilibrium fraction of late adopters.
Thus from \eqref{eq:fixed_policy_profit}, we conclude that:
\begin{equation}
\label{eq:pi_disc_d_reg2}
\rho(P_0, P_1, 0,H) = \alpha P_0 + (1-\alpha)(1-(1-\alpha)^d) P_1 .
\end{equation}

Given that for any $(P_0, P_1, 0)$  where $\alpha = 0$, $\rho(P_0, P_1, \eta,H) = 0$, while for any $(P_0, P_1, 0)$ where $0 < \alpha < 1$, it must be that $\Pi^{adopt}(\alpha, d) \geq 0$, which implies that:
\begin{equation}
\label{eq:barA_indiff}
 \bar{A} \geq P_0 - p \eta (1 - \alpha) d \geq P_0 .
\end{equation}
Thus for any pricing policy $(P_0,P_1,0)$ with $\bar{A} - P_0 < p(A_1^H - P_1)$ and $P_1 < A_1^H$, we have:
\begin{equation}
\label{eq:galpha}
 \rho(P_0, P_1, \eta,H) \leq G(\alpha) := \alpha \bar{A} + (1-\alpha)(1 - (1-\alpha)^d)A_1^H.
\end{equation}
$G$ is a continuous and concave function of $\alpha$ on the compact support $[0,1]$.  Let $\hat{\alpha}$ denote the maximizer, i.e., $\hat{\alpha} = \arg \max_{\alpha \in [0,1]} G(\alpha)$.  It is straightforward to check that under Assumption \ref{as:AlwaysFreeriding}, we must have $0 < \hat{\alpha} < 1$.  The fact that $\hat{\alpha} > 0$ is trivial.  To show $\hat{\alpha} < 1$, choose $\alpha$ such that $1 - (1 - \alpha)^d = p$, and then Assumption \ref{as:AlwaysFreeriding} ensures that $G(\alpha) > G(1) = \bar{A}$.  As shown above, for any $(P_0, P_1, \eta)$ with $\alpha = 1$, the profit is less than $\bar{A}$; thus no such policy can be optimal.

{\em Part (ii):}

This proof is straightforward. It is based on the fact that as $d \rightarrow \infty$, each agent who waits until the second period is informed with a probability
 going to 1, for any $\alpha>0$.  Second period prices can be charged at $A_1^H$ and then a price discount or referral incentive is offered that leads
 agents to be indifferent between first- and second-period adoption, which results in an equilibrium with the optimal $\hat{\alpha}^*$.
  Thus, irrespectively of whether a $\mathcal{D}$- or a $\mathcal{R}$-policy is used, only a vanishing fraction of agents need to be incentivized to adopt early as $d$ goes to
  infinity. This is enough to generate full adoption in the second stage and both limits equal $A_1^H$, the maximum allowable profit.

\end{proof}

\begin{proof}[Proposition \ref{pr:2d_limits}]

{\em Parts (i) and (ii):}

We first show that $\underset{q \rightarrow 0}{\text{lim}}  \ \underset{d_u \rightarrow \infty}{\text{lim}} \   \hat{\pi}_{\mc{R}} = \underset{q \rightarrow 0}{\text{lim}}  \ \underset{d_u \rightarrow \infty}{\text{lim}} \   \hat{\pi}$:

Let the $\mc{R}$-policy be $(P,P,\hat{\eta})$, where

\begin{equation}
\hat{\eta} = \frac{p(A^H_1 - P)(1 - (1-\tilde{f}(d_u))^{d_u}) + P - \bar{A}}{p(1-\tilde{f}(d_u))d_u}
\label{eq:eta_hat}
\end{equation}

It is easy to verify that there exist $\underline{P}$ and $\underline{d_u}$ such that for $P \in [\underline{P},A^H_1)$ and $d_u \geq \underline{d_u}$, the resulting equilibrium strategy $\mu$ is such that $\mu(d_u)=1$ and $\mu(d_l)=0$. It is thus an upper-threshold strategy where $d_u$-agents adopt early and $d_l$-agents delay adoption (the referral reward is high enough to make high-degree agents adopt early, but low enough to make low degree agents adopt late). The expression in (\ref{eq:eta_hat}) is derived from a $d_u$-agent's indifference condition $\piadopt(\alpha, d_u) = \pidefer(\alpha, d_u)$ with  $\alpha$ set to $\tilde{f}(d_u)$ (so that the information is generated only by $d_u$-agents).

Since $\tilde{f}(d_u) = \frac{q d_u}{q d_u + (1-q) d_l }$  and $\alpha = \alpha(\mu)= \tilde{f}(d_u) \mu(d_u)=\tilde{f}(d_u)$, it follows that $\underset{q \rightarrow 0}{\text{lim}}  \ \underset{d_u \rightarrow \infty}{\text{lim}} \    \alpha = 1$ and thus $\underset{q \rightarrow 0}{\text{lim}}  \ \underset{d_u \rightarrow \infty}{\text{lim}} \    \gamma_H(\mu) = 1$. Also note that $\underset{q \rightarrow 0}{\text{lim}}  \ \underset{d_u \rightarrow \infty}{\text{lim}} \    \beta(\mu) = \underset{q \rightarrow 0}{\text{lim}}  \ \underset{d_u \rightarrow \infty}{\text{lim}} \    f(d_u)  = 0$. Therefore, it immediately follows that $\underset{q \rightarrow 0}{\text{lim}}  \ \underset{d_u \rightarrow \infty}{\text{lim}} \    \rho(P,P,\hat{\eta},H) = P$. The intuition is simple: such a strategy generates, in the $q$ and $d_u$ limits, full adoption in the second period, with a vanishing fraction of early adopters who need to be incentivized. Using a policy $(A_1^H,A_1^H,\hat{\eta})$, it then follows that $\underset{q \rightarrow 0}{\text{lim}}  \ \underset{d_u \rightarrow \infty}{\text{lim}} \    \pi_{\mc{R}}(A_1^H,\hat{\eta},H)=A^H_1$ (the upper bound on profits). Thus, necessarily,  $\underset{q \rightarrow 0}{\text{lim}}  \ \underset{d_u \rightarrow \infty}{\text{lim}} \    \pi_{\mc{R}}(A_1^H,\hat{\eta},H)=\underset{q \rightarrow 0}{\text{lim}}  \ \underset{d_u \rightarrow \infty}{\text{lim}} \    \hat{\pi}_{\mc{R}} = \underset{q \rightarrow 0}{\text{lim}}  \ \underset{d_u \rightarrow \infty}{\text{lim}} \     \hat{\pi}  $.

Moreover, the policy $(A_1^H,A_1^H,\eta^+)$, where $\eta^+= \frac{ A_1^H - \bar{A}}{p(1-\tilde{f}(d_u))d_u}$ (this is just (\ref{eq:eta_hat}) with $P$ set to $A_1^H$), is optimal in the $q$ and $d_u$ limits.

We now show that $\underset{q \rightarrow 0}{\text{lim}}  \ \underset{d_u \rightarrow \infty}{\text{lim}} \    \hat{\pi}_{\mc{D}} < \underset{q \rightarrow 0}{\text{lim}}  \ \underset{d_u \rightarrow \infty}{\text{lim}} \    \hat{\pi}_{\mc{R}}$:

Let the $\mc{D}$-policy be $(P_0,P_1,0)$ for some $P_0$ and $P_1$. Any corresponding equilibrium strategy $\mu$ is a lower-threshold strategy (cf. Proposition \ref{obs:LowerThresholdMech}). If $\underset{q \rightarrow 0}{\text{lim}}  \ \underset{d_u \rightarrow \infty}{\text{lim}} \ \mu=0$, there are no early adopters and no late adopters and the profit is $0$. If  $\underset{q \rightarrow 0}{\text{lim}}  \ \underset{d_u \rightarrow \infty}{\text{lim}} \ \mu>0$, then $\underset{q \rightarrow 0}{\text{lim}}  \ \underset{d_u \rightarrow \infty}{\text{lim}} \    \beta(\mu) = \underset{q \rightarrow 0}{\text{lim}}  \ \underset{d_u \rightarrow \infty}{\text{lim}} \    (1-q)\mu(d_l) + q \mu(d_u) >0$ and thus a non-vanishing fraction of agents adopt early and pay the discounted price $P_0$. Also note that $\underset{q \rightarrow 0}{\text{lim}}  \ \underset{d_u \rightarrow \infty}{\text{lim}} \ \beta(\mu) = \mu(d_l)$ and that $\underset{q \rightarrow 0}{\text{lim}}  \ \underset{d_u \rightarrow \infty}{\text{lim}} \ \alpha(\mu) = \tilde{f}(d_l) \mu(d_l)=0$, since $\tilde{f}(d_l)$ goes to $0$. This implies that $\underset{q \rightarrow 0}{\text{lim}}  \ \underset{d_u \rightarrow \infty}{\text{lim}} \gamma_H(\mu) =0$ (there are no late adopters) since no agent can collect information. Thus, $\underset{q \rightarrow 0}{\text{lim}}  \ \underset{d_u \rightarrow \infty}{\text{lim}} \ \rho(P_0,P_0,0,H) = P_0 \beta(\mu) + P_1 \gamma_H(\mu) = P_0 \beta(\mu)$. Therefore, $\underset{q \rightarrow 0}{\text{lim}}  \ \underset{d_u \rightarrow \infty}{\text{lim}} \ \pi(P_0,P_1,0,H) \leq \bar{A} \beta(\mu) \leq \bar{A} < A^H_1$, from which it immediately follows that $\underset{q \rightarrow 0}{\text{lim}}  \ \underset{d_u \rightarrow \infty}{\text{lim}} \    \hat{\pi}_{\mc{D}}< A_1^H = \underset{q \rightarrow 0}{\text{lim}}  \ \underset{d_u \rightarrow \infty}{\text{lim}} \    \hat{\pi}_{\mc{R}}$.

Moreover, the profit-maximizing two-price policy is thus the one that gets all low-degree agents to adopt early (since they would not adopt otherwise). Since from Proposition \ref{prop:opt_two_price_pol} this can be achieved with prices $P_0=\bar{A}$ and $P_1=A_1^H$, then it follows that  $\underset{q \rightarrow 0}{\text{lim}}  \ \underset{d_u \rightarrow \infty}{\text{lim}} \ \pi(\bar{A},A_1^H,0,H) = \bar{A} \beta(\mu) = \bar{A}$ and the policy $(\bar{A},A_1^H,0)$ is an optimal two-price policy in the $q$ and $d_u$ limits.

{\em Part (iii):}

The result follows from the fact that the limiting degree distribution is $d_u$-regular and then invoking Theorem \ref{th:discount_opt_d_reg} (ii).

{\em Part (iv):}

Let $q \in (0,1)$ and let $(P,P,\eta)$ be some $\mc{R}$-policy. If a corresponding equilibrium strategy $\mu$ is such that $\underset{d_u \rightarrow \infty}{\text{lim}} \ \mu(d_u) = 0$, then since $\underset{d_u \rightarrow \infty}{\text{lim}} \tilde{f}(d_u)=1$, it follows that $\underset{d_u \rightarrow \infty}{\text{lim}} \ \alpha = \underset{d_u \rightarrow \infty}{\text{lim}} \ \mu(d_u)=0$ and thus no agent adopts late. If $\underset{d_u \rightarrow \infty}{\text{lim}} \ \mu(d_u) > 0$, then $\underset{d_u \rightarrow \infty}{\text{lim}} \ \beta(\mu) = \underset{d_u \rightarrow \infty}{\text{lim}} \ (1-q) \mu(d_l) + q \mu(d_u)>0$ and thus a non-trivial fraction of agents adopt early and have to be incentivized. From these observations, it follows that the maximal allowable profit $A^H_1$ cannot be attained. We conclude that $\underset{d_u \rightarrow \infty}{\text{lim}} \ \hat{\pi}_{\mc{R}}<A_1^H$.  
\end{proof}

\begin{proof}[Corollary \ref{cor:opt_prof_d_reg_uninformed_monop}]

{\em Part (i):}

 The monopolist's profit at a fixed policy $(P_{0,k},P_{1,k},\eta_k)$ is now defined as

\begin{equation}
\rho(P_{0,k},P_{1,k},\eta_k) =  \alpha_k (P_{0,k} - p \eta_k (1-\alpha_k)d ) + p(1-\alpha_k)(1-(1-\alpha_k)^d)P_{1,k}.
\end{equation}

 As in the proof of Theorem \ref{th:discount_opt_d_reg}, any sequence of profits associated with some sequence of $\alpha_k$'s and policies $(P_{0,k},P_{1,k},\eta_k)$ can also be achieved without referrals, i.e. by some sequence of policies $(P'_{0,k},P'_{1,k},0)$. Thus $\hat{\pi}_{\mc{D}}=\hat{\pi}$.

Note that under a $\mc{D}$-policy, an agent's indifference condition is
\begin{equation}
\label{eq:proof_corr_d_reg_indiff}
 \piadopt(\alpha_k,d) =\bar{A}-P_{0,k} = p(1 - (1-\alpha_k)^d)(A^H_1 - P_{1,k}) = \pidefer(\alpha_k,d).
\end{equation}

Now let $(P_k,P_k,\eta_k)$ be some $\mc{R}$-policy. Under such a policy, an agent's indifference condition is
\begin{equation*}
\piadopt(\alpha_k,d)= \bar{A}-P_k + p\eta_k (1-\alpha_k)d = p(1 - (1-\alpha_k)^d)(A^H_1 - P_k) = \pidefer(\alpha_k,d).
 \end{equation*}

For some $\alpha_k$ satisfying (\ref{eq:proof_corr_d_reg_indiff}), setting $P_k=P_{1,k}$ and $p \eta_k (1-\alpha_k)d - P_{1,k}=-P_{0,k}$ (or equivalently setting $\eta_k = \frac{P_{1,k} -P_{0,k}}{p (1-\alpha_k)d}$) gives us a $\mc{R}$-policy yielding the same $\alpha_k$. Moreover, doing so yields profit

\begin{eqnarray*}
\rho(P_{1,k},P_{1,k},\eta_k) &=&  \alpha_k (P_{1,k} - p \eta_k (1-\alpha_k)d ) + p(1-\alpha_k)(1-(1-\alpha_k)^d)P_{1,k} \\
 &=& \alpha_k P_{0,k} + p(1-\alpha_k)(1-(1-\alpha_k)^d)P_{1,k} \\
 &=& \rho(P_{0,k},P_{1,k},0)
\end{eqnarray*}
Thus, the same sequences of profits associated with some sequences of $\alpha_k$'s and two-price policies $(P_{0,k},P_{1,k},0)$ can also be achieved by an appropriate choice of $\mc{R}$-policy $(P_{k},P_{k},\eta_k)$.
It follows that $\hat{\pi}_{\mc{R}}=\hat{\pi}_{\mc{D}}=\hat{\pi}$, which completes the proof of Part (i).

Note that this would not work with an informed principal (as in Theorem \ref{th:discount_opt_d_reg}) because the absence of $p$ factors in the function $\rho(P_0,P_1,\eta,H)$ would lead to $\rho(P_{0,k},P_{1,k},0,H) > \rho(P_{1,k},P_{1,k},\eta_k,H) $.

{\em Part (ii):}
This proof is identical to that of Theorem \ref{th:discount_opt_d_reg}, Part (ii), except that the ex-ante probability that an agent adopts late is $p$ (the probability that $\theta=H$). It then immediately follows that the expected limiting profit equals $p A_1^H$.

\end{proof}

\section{Acknowledgments}
 We thank Francis Bloch and Itay Fainmesser for helpful discussions.  An early version of this paper was presented at the \textit{ACM Conference on Economics and Computation (EC '14).}

 Leduc gratefully acknowledges support from grant \#FA9550-12-1-0411 from the U.S. Air Force Office of Scientific Research (AFOSR) and the Defense Advanced Research Projects Agency (DARPA).

 Jackson gratefully acknowledges financial support from grant FA9550-12-1-0411
from the AFOSR and DARPA, and ARO MURI award No. W911NF-12-1-0509 and NSF grant SES-1629446.

Johari gratefully acknowledges financial support from grant FA9550-12-1-0411
from the AFOSR and DARPA.

\bibliographystyle{acmsmall}
\bibliography{acmsmall-sample-bibfile}

\end{document}